\documentclass[fleqn,usenatbib]{mnras}

\usepackage{newtxtext,newtxmath}

\usepackage[T1]{fontenc}

\DeclareRobustCommand{\VAN}[3]{#2}
\let\VANthebibliography\thebibliography
\def\thebibliography{\DeclareRobustCommand{\VAN}[3]{##3}\VANthebibliography}

\usepackage{graphicx}	
\usepackage{amsmath}	

\title[SZ hydrostatic masses with FLAMINGO]
{Relativistic SZ temperatures and hydrostatic mass bias for massive clusters in
the FLAMINGO simulations}

\author[S. T. Kay et al.]{
Scott T. Kay,$^{1}$\thanks{E-mail: scott.kay@manchester.ac.uk}
Joey Braspenning,$^{2}$
Jens Chluba,$^{1}$
John C. Helly,$^{3}$
Roi Kugel,$^{2}$
Matthieu Schaller$^{2}$
\newauthor
and Joop Schaye$^{2}$
\\
$^{1}$Jodrell Bank Centre for Astrophysics, Department of Physics and Astronomy, The University of Manchester, Oxford Road, Manchester M13 9PL, UK\\
$^{2}$Leiden Observatory, Leiden University, PO Box 9506, 2300 RA Leiden, the Netherlands\\
$^{3}$Institute for Computational Cosmology, Department of Physics, Durham University, South Road, Durham DH1 3LE, UK
}

\date{Accepted XXX. Received YYY; in original form ZZZ}

\pubyear{2015}

\begin{document}
\label{firstpage}
\pagerange{\pageref{firstpage}--\pageref{lastpage}}
\maketitle

\begin{abstract}
The relativistic Sunyaev-Zel'dovich (SZ) effect can be used to measure 
intracluster gas temperatures independently of X-ray spectroscopy. Here, 
we use the large-volume FLAMINGO simulation suite to determine whether 
SZ $y$-weighted temperatures lead to more accurate hydrostatic mass 
estimates in massive  
($M_{\rm 500c} > 7.5\times 10^{14}\,{\rm M}_{\odot}$) 
clusters than when using X-ray spectroscopic-like temperatures. 
We find this to be the case, 
on average. The median bias in the SZ mass at redshift zero is 
$\left< b \right> \equiv 1-\left< M_{\rm 500c,hse}/M_{\rm 500c,true} \right>
= -0.05 \pm 0.01$, over 4 times smaller in magnitude than the 
X-ray spectroscopic-like case, $\left< b \right> = 0.22 \pm 0.01$. 
However, the scatter in the SZ bias, $\sigma_{b} \approx 0.2$,
is around 40 per cent larger than for the X-ray case. We show 
that this difference is strongly affected by clusters 
with large pressure fluctuations, as expected from shocks in 
ongoing mergers. Selecting the clusters with 
the best-fitting generalized NFW pressure profiles, the median SZ 
bias almost vanishes,  $\left< b \right> = -0.009 \pm 0.005$, 
and the scatter is halved to $\sigma_{b} \approx 0.1$. 
We study the origin of the SZ/X-ray difference and find that,
at $R_{\rm 500c}$ and in the outskirts, SZ weighted gas better reflects the 
hot, hydrostatic atmosphere than the X-ray weighted gas. 
The SZ/X-ray temperature ratio increases with radius, a result 
we find to be insensitive to variations in baryonic physics, cosmology and 
numerical resolution. 
\end{abstract}




\section{Introduction}

A number of important cosmological tests involve measuring 
quantities that are sensitive to the growth of 
large-scale structure. The number density of massive galaxy 
clusters as a function of their mass and redshift 
is one such probe, as clusters represent the rarest, largest 
peaks in the matter density field 
\citep[e.g.][]{Allen2011,Kravtsov2012}. This method is 
particularly effective at constraining the 
parameter combination $S_{8}=\sigma_{8}(\Omega_{\rm m})^{0.5}$ 
where $\sigma_{8}$ is the linear power spectrum amplitude 
and $\Omega_{\rm m}$ the matter density parameter, but 
can also be used to constrain additional parameters such as 
the dark energy equation of state parameter, $w$ 
\citep[e.g.][]{Vikhlinin2009,Planck2016a,Pacaud2018,Chiu2023}. Complementary 
probes using clusters also yield powerful cosmological 
constraints, particularly when based on using cluster gas 
fractions in the most massive systems that retain most of their 
baryons \citep[e.g.][]{White1993,Allen2005,Mantz2014}. 

A key step in the above analyses is to calibrate cluster 
observables to mass, usually by means of a mass-observable 
scaling relation. Several methods exist for estimating cluster 
masses, namely 
weak gravitational lensing 
\citep[e.g.][for a recent review]{Umetsu2020}; 
galaxy kinematics \citep[e.g.][]{Zwicky1933,Diaferio1997,Mamon2013}
and X-ray hydrostatic analyses 
\citep[e.g.][]{Briel1992,Durret1994,Pointecouteau2005}. 
Accurate mass estimates with the latter approach requires 
high-quality X-ray data to measure radial density and 
temperature profiles of the hot intracluster medium (ICM), 
and assumes the gas is both spherically symmetric and in 
hydrostatic equilibrium. The method has been extensively 
tested with hydrodynamical simulations 
\citep[e.g.][]{Evrard1996,Kay2004,Rasia2004,Nagai2007a,Piffaretti2009,Ansarifard2020,Pearce2020}.
An important result that emerged from these theoretical studies is 
that hydrostatic masses are {\it biased}, partly due to 
incomplete thermalization of the ICM 
\citep[e.g.][]{Lau2009}
but also because the X-ray profiles themselves are biased 
towards any cooler, clumpier gas that is present
\citep[e.g.][]{Gardini2004,Mazzotta2004,Rasia2006}. The effect of 
clumping can be somewhat mitigated using techniques such as  
azimuthal filtering 
\citep[e.g.][]{Roncarelli2013,Zhuravleva2013,Eckert2015,Ansarifard2020,Towler2023}
but temperature effects are more complex 
as they result from using a single-temperature model to describe  
a multi-temperature gas \citep{Mazzotta2004}. Furthermore, 
simulations have found the spectroscopic temperature bias to be even 
more severe in the most massive clusters, leading to 
mass estimates that are biased by up to 30-40 per cent
\citep{Henson2017,Barnes2021}. 

ICM pressure profiles can also be measured using the 
thermal Sunyaev-Zel'dovich (tSZ) 
effect (e.g. \citealt{Sunyaev1972,Birkinshaw1999};
for a recent review, see \citealt{Mroczkowski2019}). 
The tSZ signal results from the CMB photons undergoing inverse 
Compton scattering off the (more energetic) thermalised 
electrons in the ICM, leading to a boost in photon energy 
that distorts the CMB blackbody spectrum on $\sim$arcminute 
angular scales. It was first measured as a CMB decrement 
(reduction in intensity) in the 1970s using single dish radio 
telescopes at $\sim 10{\rm GHz}$ 
\citep[e.g.][]{Pariiskii1972,Gull1976}. Four decades later, 
hundreds to thousands of SZ clusters have been detected 
in CMB surveys such as the space-based {\it Planck} 
satellite \citep{Planck2016b} and ground-based facilities 
such as the South Pole Telescope \citep[SPT;][]{Bleem2015} 
and the Atacama Cosmology Telescope \citep[ACT;][]{Hilton2021}.
Many SZ studies are now yielding group and cluster pressure profiles
\cite[e.g.][]{Aslanbeigi2013,Planck2013,Sayers2013,Bourdin2017,Pratt2021}, 
often by combining data from more than one telescope to probe 
a wider range of spatial scales
\citep[e.g.][]{Ruppin2018,Perrott2019,Pointecouteau2021,Melin2023,Oppizzi2023}.
Many of these studies show the pressure profile is reasonably 
{\it universal}, following a generalised Navarro, Frenk and White 
model (GNFW; \citealt{Nagai2007b}), with model parameters similar 
to those suggested from the analysis of an X-ray sample by 
\citet{Arnaud2010}. 

Hydrostatic masses cannot be estimated with pressure profiles 
alone but the combination of SZ (pressure) and X-ray (density) 
data allows this to be achieved without expensive X-ray spectroscopy
\citep[e.g.][]{Ameglio2009,Tchernin2016,Eckert2019}.
An alternative possibility that bypasses X-ray data completely, 
is to measure the relativistic SZ cluster signal.
In hotter clusters ($T>5\,{\rm keV}$ or so), relativistic effects are more important and affect the spectral distortion of the tSZ effect
\citep[e.g][]{Rephaeli1995,Challinor1998,Itoh1998,Sazonov1998,Chluba2012}.
This overall {\it correction} can 
be modelled as a function of the electron temperature so, in principle, can be used to 
determine the temperature of the ICM. Such measurements will 
also be significant for cluster astrophysics as it will 
provide measurements of cluster temperatures independent of X-ray observations. 
\citep[e.g.][]{Pointecouteau1998,Hansen2004,Chluba2012,Chluba2013}.
Previous simulation-based studies have shown that the 
Compton-$y$ weighted 
temperature is a low-scatter mass proxy and less sensitive to 
cluster physics than the X-ray temperature 
\citep{Kay2008,Lee2020,Lee2022}. Furthermore, neglecting the temperature 
dependence of the tSZ signal can also bias the tSZ flux ($Y$) 
and, in turn, lead to biases in cosmological parameters 
\citep{Remazeilles2019,Rotti2021}.

A logical follow-on question, 
the subject of this paper, is whether hydrostatic 
mass estimates using SZ temperatures (and thus SZ-only data) 
are less biased than X-ray masses. We show that this is indeed 
the case (a lower average bias) but with an interesting caveat: 
the scatter increases due to the clusters undergoing major mergers, 
an effect that becomes more prominent at higher redshift.
We also show that the gas that contributes most to the 
$y$-weighted temperature (i.e. the gas with the highest pressure) 
is more diffuse and {\it hydrostatic} (i.e. has lower bulk/infall  
velocity), on average, than the gas with the highest X-ray emissivity, 
especially in the cluster outskirts. 
Our analysis uses the new FLAMINGO suite of 
large-volume hydrodynamic simulations \citep{Schaye2023,Kugel2023} with 
the flagship, 2.8 Gpc box containing hundreds of massive 
($>10^{15}\,{\rm M}_{\odot}$) clusters that are the  
most suitable for measuring the relativistic SZ effect. We also make 
use of the smaller (1 Gpc) boxes to investigate the sensitivity 
of our results to changes in the resolution, gas physics 
and cosmological model. This study complements the work by 
Braspenning et al. (in preparation) who use the same FLAMINGO 
dataset to study hydrostatic mass bias in comparison with X-ray observations.

The rest of the paper is organised as follows. 
In Section~\ref{sec:method} we summarise details of the FLAMINGO 
simulations, introduce the key equations of the relativistic tSZ 
signal used here and how we estimate these, and other relevant 
properties, from the simulations. Our main results are then broken 
into 2 sections. In Section~\ref{sec:global}, we present global SZ 
temperature-mass relations and how these compare with other 
temperatures in the FLAMINGO simulations. Then, in 
Section~\ref{sec:profmass}, we focus on the radial gas pressure 
and temperature profiles, and hydrostatic masses, as well as 
looking at the properties of the SZ $y$-weighted gas in more 
detail. Our results are summarized and conclusions drawn 
in Section~\ref{sec:conclusions}.

\section{Theory and Simulations}
\label{sec:method}

In this section, we outline the relativistic tSZ effect and 
how it can be used to measure the Compton-$y$ weighted 
temperature of the ICM.  We then describe the FLAMINGO simulations 
before discussing how the SZ (and other relevant) properties 
are calculated from the particle data.

\subsection{Relativistic thermal Sunyaev-Zel'dovich effect}
\label{sec:relSZ}

The relativistic tSZ effect produces a change in the observed 
intensity of the CMB radiation at frequency $\nu$ along the 
line of sight
\begin{equation}
    {\Delta I_{\nu} \over I_{0}} = f(\nu,T_{\rm e}) \, y,
\end{equation}
where $I_{0}=2(k_{\rm B}T_{\rm CMB})^{3}/(hc)^{2}$ and 
$T_{\rm CMB}=2.725\,{\rm K}$ is the mean CMB temperature.
The function $f(\nu,T_{\rm e})$ describes the shape of the 
relativistic spectral distortion from a 
thermalized electron gas with temperature $T_{\rm e}$
\cite[e.g.][]{Chluba2012,Chluba2013,Remazeilles2019}.
\footnote{Note that $f(\nu,0)$ is the spectral distortion shape in the 
non-relativistic limit, as is normally assumed in most current tSZ analyses.}
The amplitude of the tSZ distortion along a given line of sight 
is determined by the Compton-$y$ parameter, an integral of the 
electron thermal pressure, $P_{\rm e}$, as
\begin{equation}
    y = \frac{\sigma_{\rm T}}{m_{\rm e}c^2} \, \int P_{\rm e} \, {\rm d}l
    = \int \tilde{y} \, {\rm d}l,
\end{equation}
where $\tilde{y} \propto P_{\rm e}$ is the contribution to $y$ 
per unit length along the line of sight. The tSZ flux density 
from a given solid angle of sky, $\Omega$, is given by 
\begin{equation}
    S_{\nu} = I_{0} \int_{\Omega} f(\nu,T_{\rm e}) \, y \, {\rm d}\Omega 
    = I_{0} d_{\rm A}^{2}(z) 
    \int_{V} f(\nu,T_{\rm e}) \, \tilde{y} \, {\rm d}V,
\label{eqn:SZflux}
\end{equation}
assuming, for the second equality, all the electrons are at the same redshift  
$z$ and angular diameter distance $d_{\rm A}$. 
For an isothermal gas, this equation simplifies to 
\begin{equation}
S_{\nu}=I_{0} d_{\rm A}^{2}(z) \, f(\nu,T_{\rm e})\,Y,
\label{eqn:SZflux_isothermal}
\end{equation} 
where we have defined
\begin{equation}
    Y = \int \tilde{y} \, {\rm dV},
\end{equation}
a quantity that is often also referred to as the tSZ flux 
and is proportional to the 
integrated thermal energy of the electrons. In this case, both 
the electron temperature ($T_{\rm e}$) and flux ($Y$) can be 
simultaneously measured from multi-frequency CMB data. 

In practice, the ICM is not isothermal. Clusters are known to have 
declining temperature profiles beyond the core 
\citep[e.g.][]{Markevitch1998,Vikhlinin2005}.
Simulations also predict the gas to have a range of temperatures within 
each radial shell 
\citep[e.g.][]{Lee2020,Barnes2021}.
The measured temperature from tSZ data will therefore be a weighted 
average over the gas volume. To account for this, we follow the 
approach of \cite{Chluba2013}, writing the spectral shape, 
$f(\nu,T_{\rm e})$ as a Taylor expansion about a pivot temperature $T_{0}$
\begin{equation}
    f(\nu,T_{\rm e}) = 
    f(\nu,T_{0}) + 
    \partial f \Delta T_{\rm e} + 
    \tfrac{1}{2} \partial^{2} \! f (\Delta T_{\rm e})^{2} + \ldots,
\end{equation}
where $\Delta T_{\rm e} = T_{\rm e}-T_{0}$ and 
\begin{equation}
    \partial^{k}\! f = 
    \left. \frac{\partial^{k}\! f}{\partial T_{\rm e}^{k}} 
    \right|_{T_{0}}.
\end{equation}
This allows us to write the volume integral in equation~\ref{eqn:SZflux} 
for the flux as
\begin{equation}
\int \! \tilde{y} f(\nu,T_{\rm e}){\rm d}V 
\approx 
f(\nu,T_{0})Y + 
\partial f \! \int \! \tilde{y} \Delta T_{\rm e}{\rm dV},
\label{eqn:tSZseries}
\end{equation}
to first order accuracy. Defining the Compton-$y$-weighted temperature as
\begin{equation}
    T_{y} = {1 \over Y} \int \tilde{y} \, T_{\rm e} \, {\rm d}V,
\label{eqn:Ty}
\end{equation}
the first-order term in equation~\ref{eqn:tSZseries} will 
vanish if we set $T_{0}=T_{y}$. We can thus use 
equation~\ref{eqn:SZflux_isothermal}, replacing 
$T_{\rm e}$ with $T_{y}$, to calculate the flux density, 
with a best-fit solution yielding values for $\{ Y, T_{y} \}$.
In this paper, we will focus on the effect of using $T_{y}$ 
profiles to measure hydrostatic masses but we also provide results for 
the second, third and fourth order temperature moments, required 
for more accurate relativistic flux calculations, in 
Appendix~\ref{sec:res_higher}.

\subsection{The FLAMINGO simulations}

We model the relativistic tSZ signal from clusters using the 
FLAMINGO simulations. These are a set of large-volume (Gpc-scale)
cosmological simulations that follow the growth of large-scale 
structure in the dark matter, baryon and neutrino components.
Full details of the simulations including comparisons with 
key observational data and the model calibration process can 
be found in \cite{Schaye2023} and \cite{Kugel2023} respectively.
The FLAMINGO simulations are especially useful for this work, 
for several reasons. 
Firstly, they contain a large number of massive 
($M \sim 10^{15}\,{\rm M}_{\odot}$) clusters, the objects expected to 
produce the largest relativistic tSZ signal. 
Secondly, the fiducial model is calibrated to reproduce two key 
observables (the $z \approx 0$ galaxy stellar mass function and 
cluster X-ray hot 
gas fractions at $z \approx 0.1$). This model predicts cluster scaling 
relations and thermodynamic profiles that agree well with X-ray and SZ 
cluster observations, even though the halo mass range of the gas 
fraction calibration 
($10^{13.5}<M_{\rm 500c}/{\rm M}_{\odot}<10^{14.36}$ for fiducial 
resolution; \citealt{Kugel2023}) 
is outside that of massive systems \citep{Schaye2023,Braspenning2023}. 
Thirdly, there are a suite of large-volume runs with varying subgrid models 
and cosmological parameters, allowing us to assess the robustness 
of our results to such variations in the astrophysical and 
cosmological parameter space.

The FLAMINGO simulations assume a default, spatially flat $\Lambda$CDM 
cosmology with parameter values taken from the Dark Energy Survey year 3 analysis 
including external constraints
(the '3$\times$2pt + All Ext.' model). The key values are: 
$\Omega_{\rm m}=0.306; \Omega_{\rm b}=0.0486; \Omega_{\nu}=0.00139; 
h=0.681; \sigma_{8}=0.807$, with the sum of the neutrino masses set to 
$\sum m_{\nu} c^{2}= 0.06\,{\rm eV}$. 
Initial conditions were created using a modified version of 
{\sc monofonic} \citep{Hahn2021} that includes neutrinos using the method 
described in \cite{Elbers2022}. This assumes third order Lagrangian perturbation theory and uses separate transfer functions to generate the dark matter, baryon 
and neutrino perturbations. The random phases 
for the Fourier modes were generated using {\sc panphasia} (first described in 
\citealt{Jenkins2013}).

All simulations were run using the {\sc swift} $N$-body/hydrodynamics code 
\citep{Schaller2023}. Gravitational forces are calculated using the 
Particle-Mesh algorithm on large scales and the fast multipole method 
on small scales. Hydrodynamical forces are calculated for gas 
particles using the SPHENIX implementation of the 
Smoothed Particle Hydrodynamics (SPH) method 
\citep{Borrow2022}. SPHENIX is a density-energy SPH scheme that 
incorporates both artificial viscosity and artificial conduction terms, as well 
as using a higher-order (Wendland C2) smoothing kernel. Massive neutrinos 
are evolved as a separate particle species using the $\delta f$ method 
of \citet{Elbers2021}.

Gas radiative cooling and heating rates are implemented using the 
method described in \citet{Ploeckinger2020} that use tabulated 
rates from {\sc cloudy} \citep{Ferland2017}. Hydrogen and helium 
reionization occur at redshifts $z=7.8$ and $z=3.5$ respectively.
Gas with hydrogen density $n_{\rm H}>10^{-4}\,{\rm cm}^{-3}$ and 
overdensity $\delta>10$ is forced to have a minimum pressure 
$P_{\rm min} \propto n_{\rm H}^{4/3}$ to reflect an unresolved 
multiphase interstellar medium. As discussed in \citet{Schaye2023}, 
this pressure floor corresponds to a constant Jeans mass of 
$M_{\rm J} \sim 10^{7}\,{\rm M}_{\odot}$ but is unresolved in the 
FLAMINGO simulations. 

Star formation is modelled using the method described in 
\cite{Schaye2008}. Gas particles with $\delta>100$, $n_{\rm H}>n_{\rm H}^{*}$
(with $n_{\rm H}^{*}=0.1\,{\rm cm}^{-3}$) and $1<P/P_{\rm min}<2$ 
are stochastically converted into collisionless star particles at a 
pressure-dependent rate that matches the observed Kennicutt-Schmdt law
\citep{Kennicutt1998}. Stellar mass loss from 
stellar winds, AGB stars, Type~Ia and core-collapse supernovae (SNe) is 
modelled through mass transfer from the star particle to 
surrounding gas particles \citep{Wiersma2009,Schaye2015}.
Nine elements are tracked separately (H,He,C,N,O,Ne,Mg,Si and Fe). 

Supernova feedback primarily comes from
the core-collapse SNe, assuming an available specific energy of 
$1.18 \times 10^{49}f_{\rm SN} \, {\rm erg}{\rm M}_{\odot}^{-1}$.
SN energy is added in kinetic form, kicking opposing pairs of particles 
with a wind speed $\Delta v_{\rm SN}$ \citep{Chaikin2022}. 
Black hole growth is modelled using 
the modified version or the Bondi-Hoyle accretion rate by 
\citet{Booth2009}, capped at the Eddington rate. This uses a density 
dependent boost factor, $\alpha = (n_{\rm H}/n_{\rm H}^{*})^{\beta_{\rm BH}}$, 
to account for numerical (unresolved Bondi radius) and physical (single 
phase ISM) limitations in the simulations. 
As in \cite{Booth2009}, AGN feedback is included by 
raising the temperature of the nearest gas particle by $\Delta T_{\rm AGN}$, 
once sufficient mass has been accreted by the black hole and 1.5 per cent of this 
mass is available for heating. The subgrid parameters used for 
calibration, are:
$\{ f_{\rm SN}, \Delta v_{\rm SN}, \Delta T_{\rm AGN}, \beta_{\rm BH} \}$.

\begin{table}
	\centering
	\caption{Simulations with varying physics. Column~1 gives the run label 
 and column~2 briefly describes the key differences from the fiducial case.
 All runs adopt the same box-size and particle numbers as L1\_m9.}
	\label{tab:flamingo_models}
	\begin{tabular}{ll} 
		\hline
        Label & Brief Description\\
		\hline
        L1\_m9 & Fiducial calibration\\ 
        fgas$+2\sigma$ & $2\sigma$ higher gas fractions\\  
        fgas$-2\sigma$ & $2\sigma$ lower gas fractions\\ 
        fgas$-4\sigma$ & $4\sigma$ lower gas fractions\\ 
        fgas$-8\sigma$ & $8\sigma$ lower gas fractions\\ 
        Jet            & Jet feedback\\ 
        Jet\_fgas$-4\sigma$ & Jets with $4\sigma$ lower gas fractions\\ 
        M*-1$\sigma$   & $1\sigma$ lower stellar masses\\ 
        M*-1$\sigma$\_fgas$-4\sigma$ & $1\sigma$ lower stellar masses, $4\sigma$ lower gas fractions \\ 
        Planck          & Planck cosmology\\
        PlanckNu0p24Var & Higher neutrino mass, varying cosmo parameters\\
        PlanckNu0p24Fix & Higher neutrino mass, fixed cosmo parameters\\
        LS8             & Lower power spectrum amplitude\\
		\hline
	\end{tabular}
\end{table}

The largest FLAMINGO hydrodynamics run is labelled L2p8\_m9 and will be 
the main simulation that is analysed here. This run contains $5040^{3}$ gas 
and dark matter particles each, and $2800^{3}$ neutrino particles, 
within a box size of 2.8 comoving Gpc. The gas particle mass for this run 
is $m_{\rm gas}=1.07 \times 10^{9}\,{\rm M}_{\odot}$ and the maximum 
physical softening length is $\epsilon_{\rm max}=5.7\,{\rm kpc}$. This 
run has subgrid physics parameters calibrated to match the low redshift 
galaxy stellar mass function and cluster gas fractions, achieved using 
a machine learning emulator-based approach \citep{Kugel2023}. The values of 
the key subgrid parameters for this model are: 
$\{ f_{\rm SN},\Delta v_{\rm SN},\Delta T_{\rm AGN},\beta_{\rm BH} \} = 
\{ 0.238, 562\,{\rm kms}^{-1}, 10^{7.95}\,{\rm K}, 0.514 \}$.

We also make use of the suite of 1 Gpc (L1) runs, the fiducial calibrated case 
referred to as L1\_m9. The other runs include variations to the 
subgrid parameters to produce higher/lower cluster gas fractions and 
stellar masses, offset by a fixed number of observed standard deviations; 
runs with jet feedback using the model of \citet{Husko2022}, 
and runs with varying cosmological 
parameters and assumptions for the neutrino species. 
Note that lower gas fractions are mainly the result of stronger AGN feedback 
(higher heating temperature or jet speed) while lower stellar masses are 
mainly from stronger SN feedback (energy fraction and wind speed). 
The L1 runs used in this paper are summarised in 
Table~\ref{tab:flamingo_models}; see \citet{Schaye2023} for full details, 
with their table 1 listing subgrid parameter values for each run.

\subsection{Simulated cluster properties}

Dark matter haloes are identified in the FLAMINGO snapshot 
data using the VELOCI{\sc raptor} phase space-based 
halo finder \citep{Elahi2019}. 
This code defines the halo centre as the particle with the lowest binding energy 
(referred to as the centre of potential) and separates bound particles 
into a central object and its satellites. A second code, 
Spherical Overdensity and Aperture Processor ({\sc soap}) is then run, 
to calculate various halo properties within a range of apertures. 
For this paper, we define the halo's mass and radius such that the 
mean density of the sphere, centred on the halo, is 
$\left< \rho \right>=500\,\rho_{\rm cr}$ 
where $\rho_{\rm cr}(z)=3H^2/8\pi G$ is the critical density. 
The Hubble parameter 
$H(z) \approx \sqrt{\Omega_{\rm m}(1+z)^{3}+1-\Omega_{\rm m}}$ 
as we only consider flat models with a cosmological constant and 
neglect, here, the subdominant contribution from photons and 
neutrinos at low redshifts, as appropriate for this paper. 
The corresponding mass of the halo is 
\begin{equation}
    M_{\rm 500c} = \tfrac{4}{3}\pi R_{\rm 500c}^{3} \, 
    500\, \rho_{\rm cr}(z).
\end{equation}

Our main results are volume-integrated averages of the halo's gas properties, 
either from within $R_{\rm 500c}$ (scaling relations) or from spherical shells 
(3D radial profiles).
\footnote{We do not exclude gas in substructures from our analysis.}
In general, these averages can be written, for a continuous distribution, as 
\begin{equation}
    \left< A \right> = \frac{1}{W} \int_{V} w A \, {\rm d}V,
\end{equation}
where $w$ is the weight, $A$ is the property being averaged and the normalisation 
constant is 
\begin{equation}
    W = \int_{V} w\,{\rm d}V.
\end{equation}
For $y$-weighted averages, as discussed above, we can define 
$w = n_{\rm e}T$, assuming $T=T_{\rm e}$ (i.e. electrons and ions 
in the volume element ${\rm d}V$ have equal thermal energies). 
We also use volume weighting ($w=1$),  
mass weighting ($w=\rho$) and spectroscopic-like weighting 
($w=\rho^{2}T^{-3/4}$), where $\rho$ is the hot 
gas mass density. 
Volume weighting is used for electron densities and pressures whereas
mass weighting is used for temperatures and velocities 
(note that $Y$ is proportional to 
the mass-weighted temperature). Spectroscopic-like weighting is used 
as a proxy for X-ray spectroscopic temperatures of hot 
($k_{\rm B}T>2\,{\rm keV}$) clusters where the X-ray emission is 
dominated by thermal bremmstrahlung \citep{Mazzotta2004}.

As we are analysing data from SPH simulations, the above integrals 
must be approximated as discrete sums over the hot gas particles, 
each with effective volume $m/\rho$, where $m$ is the particle's mass 
and $\rho$ its SPH density. Thus, the discrete average becomes 
\begin{equation}
    \left< A \right> = 
    \frac{1}{W}
    \sum_{i=1}^{N} w_{i} A_{i} \left( m_{i}/\rho_{i} \right),
\end{equation}
for all $N$ gas particles with temperature $T_{i}>T_{\rm min}$ in volume $V$.
For $y$-weighted properties we can set $T_{\rm min}=0$ since 
$n_{\rm e}=0$ for neutral gas by definition (these values are calculated 
for every gas particle in the FLAMINGO simulations). For 
spectroscopic-like properties, we set $T_{\rm min}=10^{6}\,{\rm K}$ 
($k_{\rm B}T_{\rm min} \sim 0.1\,{\rm keV}$) 
for consistency with previous work. For volume- and 
mass-weighted properties, we also set $T_{\rm min}=10^{6}\,{\rm K}$ 
when considering radial profiles of massive clusters but reduce this 
by an order of magnitude when 
including lower mass groups in our scaling relations. 

For each halo, we exclude gas particles that were recently heated 
by AGN (within the past $15\,{\rm Myr}$) as these particles are 
briefly very hot and dense. (In practice, this cut is 
not important as it affects a very small fraction of the particles.)
In the case of global spectroscopic-like weighted temperatures, we 
also exclude gas particles 
from within the core ($r<0.15\,R_{\rm 500c}$) as this 
temperature can be significantly affected by the presence of 
cooler, denser particles in this region 
(X-ray observations also show more temperature scatter here).
Furthermore, the models are less likely to be reliable on 
these scales where the physics is more complex. 

We scale each cluster property with mass and redshift using the expected 
{\it self-similar} scalings from gravitational structure formation 
\citep[e.g.][]{Voit2005}. These functions, for the temperature, 
electron pressure and velocity components respectively, 
are as follows
\begin{eqnarray}
    T_{\rm 500c} &=& 
    \frac{\mu m_{\rm p}}{k_{\rm B}} \, \frac{GM_{\rm 500c}}{2R_{\rm 500c}} \\
    P_{\rm 500c} &=&
    \frac{500\,f_{\rm b}\,\rho_{\rm cr}}{\mu_{\rm e}m_{\rm p}}\,k_{\rm B}T_{\rm 500c} \\
    v_{\rm 500c} &=& 
    \sqrt{\frac{GM_{\rm 500c}}{R_{\rm 500c}}},
\end{eqnarray}
where the last case is just the circular velocity of the halo at $R_{\rm 500c}$. 
We set the cosmological baryon fraction, 
$f_{\rm b}=\Omega_{\rm b}/\Omega_{\rm m}\approx 0.16$, mean atomic weight 
$\mu=0.59$ and mean atomic weight per electron $\mu_{\rm e}=1.14$. 

\section{Global SZ Temperatures}
\label{sec:global}

We first assess how the global (halo-averaged) SZ 
($y$-weighted) temperature, $T_{y}$, compares 
with the mass-weighted temperature, $T_{\rm m}$, and the 
spectroscopic-like temperature, $T_{\rm sl}$. 
While our main results will focus on massive clusters 
($M_{\rm 500c}>7.5 \times 10^{14}\,{\rm M}_{\odot}$) only, here 
we extend the mass range down to low-mass groups 
$M_{\rm 500c} > 10^{13}\,{\rm M}_{\odot}$), covering 
around two orders of magnitude in halo mass. Groups 
are more sensitive to non-gravitational physics than 
clusters, with feedback heating and ejecting more 
gas, leading to lower baryon fractions.
As a result, we are likely to see 
larger variations between the temperature measures 
in lower-mass objects. Furthermore, calibrating 
the temperature-mass relation down to group scales 
might be useful for constructing statistical 
predictions (e.g. stacked halo measurements). 

In addition to the above mass limit, we impose a 
further constraint that the hot ($T>10^{5}\,{\rm K}$) 
gas mass in the halo must be 
$M_{\rm gas,500c}>5\times 10^{10}\,{\rm M}_{\odot}$ 
(around 50 gas particles at the fiducial resolution).
This only affects a small number of gas-poor groups 
close to the total mass limit where temperatures 
could not be reliably defined.
 
\subsection{The LLR method}

One of the consequences of cooling and feedback 
effects is that mass-observable scaling relations 
can no longer be accurately described using a single 
power law over the mass range of groups and 
clusters i.e. they are no longer self-similar.
We instead model the temperature-mass relations 
adopting the approach of \citet{Farahi2018} who 
use the local linear regression (LLR) method. 
The LLR method produces local properties 
(e.g. normalisation, slope and scatter) 
within each halo mass bin. 
Defining $s=\ln T$ and $\mu=\ln M$, the 
expectation value of $s$ at fixed $\mu$ (i.e. the 
temperature-mass relation) is modelled as a 
linear function in the vicinity of $\mu$
\begin{equation}
\left< s|\mu \right> = \pi(\mu) + \alpha(\mu) \mu,
\end{equation}
where the normalisation, $\pi$ and slope, $\alpha$ 
are expected to vary smoothly with halo mass 
(and would be constant in the case of a pure power 
law). The best-fitting $(\pi,\alpha)$ 
values are calculated at fixed $\mu$ by minimizing 
the function 
\begin{equation}
\epsilon^{2}(\mu) = \sum_{i=1}^{N} \, w_{i}^{2} \, 
\left[] s_{i}-\pi-\alpha \mu_{i}(\mu)\right]^{2},
\end{equation}
where the index $i$ in the sum runs over the $N$ haloes 
in the sample, each with 
$\mu_{i}(\mu) =\ln(M_{i})-\mu=\ln(M_{i}/M)$. 
The weight, $w_i$ is defined as a Gaussian 
function, centred on $\mu$
\begin{equation}
w_i = {1 \over \sqrt{2\pi}\sigma_{\rm LLR}} \, 
\exp \left( - {\mu_{i}^{2} \over 
2\sigma_{\rm LLR}^{2}} \right),
\end{equation}
where $\sigma_{\rm LLR}=0.46$ (0.2 dex in $M$). 
This ensures that the slope and normalisation 
are primarily determined by haloes with 
similar mass (i.e. the {\it local} in LLR). 

The covariance matrix for two variables, $a$ and $b$, is then 
estimated as
\begin{equation}
    C_{ab} = A \, \sum_{i=1}^{N} w_{i} \, \delta s_{a,i} \delta s_{b,i},
\end{equation}
where $\delta s_{i}=s_{i}-\pi-\alpha \mu_{i}$ is 
the deviation of $s$ from the LLR model for $\mu$. 
(Note that each variable has their own set of LLR 
fit parameters e.g. $\pi_{a}(\mu)$ and 
$\alpha_{a}(\mu)$.) The normalisation constant $A$ 
is defined as 
\begin{equation}
    A = {\sum_{i=1}^{N} w_{i} \over \left( 
    \sum_{i=1}^{N} w_{i} \right)^{2} - 
    \sum_{i=1}^{N} w_{i}^{2}},
\end{equation}
which results in $C_{ab}$ being an unbiased estimator 
for the covariance matrix. The diagonal terms yield 
the local scatter in each variable at fixed mass
\begin{equation}
\sigma_{a} = \sqrt{A \, \sum_{i=1}^{N} w_{i} \, 
\delta s_{a,i}^{2}}.
\label{eqn:scatter}
\end{equation}

We calculate the LLR parameters $(\pi,\alpha,\sigma)$ 
for each scaling relation at 50 equally-spaced 
$\mu$ values in the mass range 
$10^{13}<M_{\rm 500c}/{\rm M}_{\odot}<10^{15.5}$. 
We then discard results for $\mu$ values 
within $\sigma_{\rm LLR}$ of these mass limits, 
and where there are fewer than 10 objects in the range 
$\mu \pm \Delta \mu/2$ where 
$\Delta \mu \approx 0.115$ 
(the spacing between adjacent $\mu$ values). 
Note this latter constraint only affects the 
largest $\mu$ values/halo masses. 
We checked that these criteria also lead to converged results 
for the smaller-box L1 runs at $z=0$, used below.

\subsection{Fiducial model}

\begin{figure*}
\centering
\includegraphics[scale=0.6]
{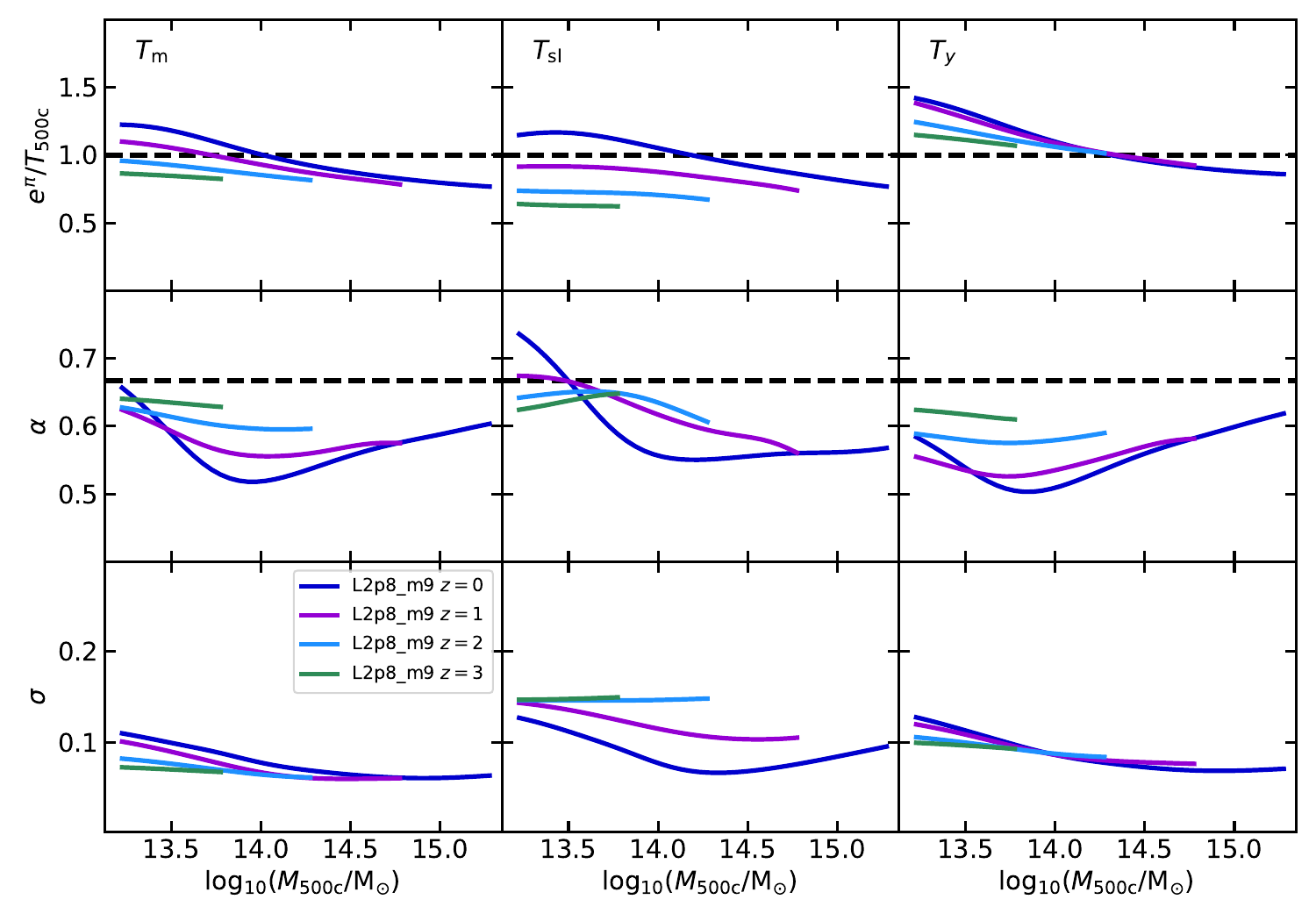}
\caption{LLR parameters for the 3 temperature-mass 
relations at varying redshift ($z=0,1,2,3$) for 
the fiducial run L2p8\_m9. From top to bottom, the 
rows show the local normalisation ($e^{\pi}$), 
slope ($\alpha$) and scatter ($\sigma$) of the 
temperature as a function of halo mass, 
$M_{\rm 500c}/{\rm M}_{\odot} \equiv e^{\mu}$. 
Each column shows results for the different 
temperatures ($T_{\rm m}$, $T_{\rm sl}$ and $T_{y}$ 
respectively). Note that the normalisation is shown 
relative to the gravitational temperature 
$T_{\rm 500c}$ (the dashed horizontal line is where 
$e^{\pi}=T_{\rm 500c}$). The horizontal dashed line 
in the middle row is the self-similar slope, 
$\alpha=2/3$.
}
\label{fig:tmrel_z_llr}
\end{figure*}

We first look at the LLR parameters for the 
fiducial L2p8\_m9 run, shown as a function of 
halo mass at redshifts $z=0,1,2,3$ in Fig.~\ref{fig:tmrel_z_llr}. 
The temperature normalisation, $\exp(\pi)$, results are shown in the 
top panels. We plot this relative to $T_{\rm 500c}$ for each halo 
mass bin to highlight the differences between results and to remove 
the self-similar redshift dependence 
($T \propto E(z)^{2/3}$ at fixed mass). For all 3 temperature 
weightings at $z=0$ (dark blue curves), we see that 
$T>T_{\rm 500c}$ on group scales 
($M_{\rm 500c}<10^{14}\,{\rm M}_{\odot}$) but 
$T<T_{\rm 500c}$ for the more massive clusters. Consequently, 
in the middle panels we see that the local slope, $\alpha$, is smaller  
than the self-similar scaling (2/3) except for the 
spectroscopic-like temperature at the lowest masses, where this is 
no longer a reliable X-ray proxy. On group scales, the slope 
becomes flatter with increasing mass, reaches a minimum around 
$M_{\rm 500c} \sim 10^{14}\,{\rm M}_{\odot}$, then starts increasing 
again but never quite reaches the self-similar value in massive 
clusters (we will discuss this further, below). 
The scatter (bottom panels) is typically quite low 
($\sigma \sim 0.05-0.1$) and is largest in the low-mass groups.
\footnote{
We checked the temperature scatter distribution on cluster scales, 
$10^{14}<M_{\rm 500c}/{\rm M}_{\odot}<10^{15}$, and it 
is close to log-normal for all 3 temperature weightings.
}
This mass dependence is expected from non-gravitational 
processes (radiative cooling and subsequent feedback) that 
result in gas with higher entropy (hotter and less dense) 
in groups than in clusters. 

Comparing the 3 temperature weightings (different columns), 
the SZ $y$-weighted 
temperature is higher in groups than the mass and 
spectroscopic-like values, but closest to the gravitational 
temperature in massive clusters. The first point means that the gas 
with the highest pressure in groups is hotter, as can be 
expected from the effects of thermal feedback, heating the densest 
gas and ejecting it from the halo, a process that is more effective 
in lower mass objects with shallower gravitational potentials. 
On the other hand, the spectroscopic-like 
temperature is most affected by cooler (but still hotter than 
0.1 keV), denser gas. Such gas becomes more prevalent in more massive 
clusters which likely explains why the slope increases more gradually 
with mass on cluster scales than for the other temperature measures 
(e.g. see \citealt{Barnes2017}).

The results at higher redshift span a narrower range in mass 
due to the paucity of high-mass objects there. However, the 
2.8 Gpc box is still sufficiently large to contain reasonable 
numbers ($>10$) of 
clusters ($M_{\rm 500c}>10^{14}\,{\rm M}_{\odot}$) at $z=2$ 
and 
high-mass groups ($10^{13.5}<M_{\rm 500c}/{\rm M}_{\odot}<10^{14}$) at $z=3$, 
making it useful for predicting the properties of high 
redshift objects in future deep cluster surveys. In all 3 
temperature cases, looking at group scales, the 
normalisation is lower at higher redshift, i.e. the 
evolution is {\it slower} than self-similar, or, at 
fixed mass, objects at higher redshift are colder than 
expected from gravitational heating. 
The SZ temperature normalisation evolves closest to the 
self-similar rate (and is almost perfectly self-similar on cluster 
scales at $z<2$) whereas the spectroscopic-like temperature is the 
least self-similar. 
The slope moves closer to the self-similar value at higher redshift 
while the scatter is almost constant with redshift in the 
mass-weighted and $y$-weighted cases (spectroscopic-like 
temperatures show more scatter at higher redshift but again, 
these results are not reliable on group scales).

Our results are qualitatively consistent with those found by 
\cite{Lee2022} who analysed temperature-mass relations for the 3
weightings applied to 4 different simulation sets at $z=0-1.5$. 
Temperature-mass relations from FLAMINGO simulations at $z=0-2$ are also 
presented in \cite{Braspenning2023}. There, the median 
mass-weighted temperature is plotted against halo mass and 
the results are shown to be in good agreement with X-ray observations 
at $z<0.6$.

\subsection{Other models}

\begin{figure*}
\centering
\includegraphics[scale=0.6]{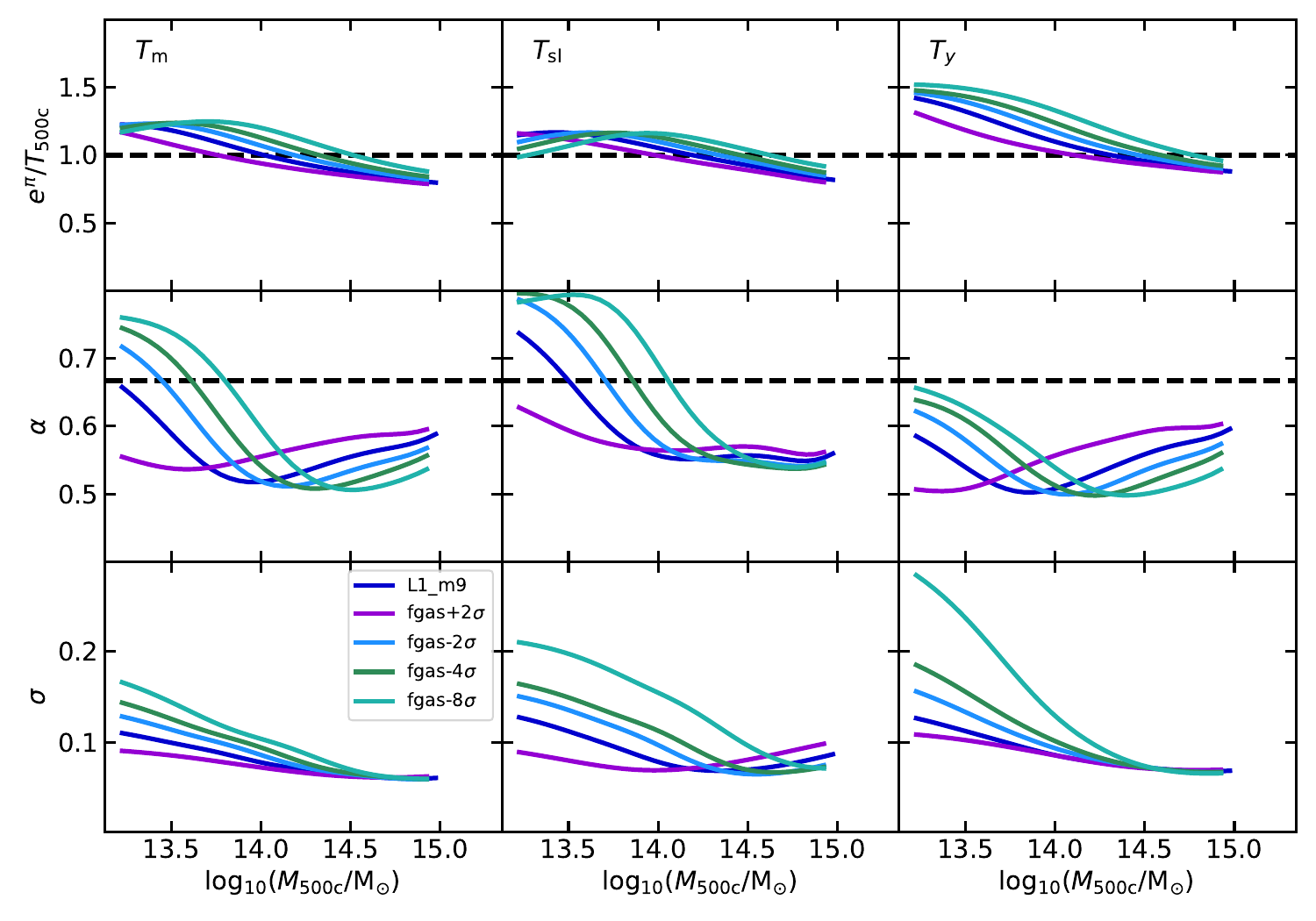}
\caption{Temperature-mass relations for runs that are calibrated 
to shifted observed cluster hot gas fractions (by the indicated 
number of $\sigma$ in the run label). 
Details are as in Fig.~\ref{fig:tmrel_z_llr}.}
\label{fig:tmrel_agn_llr}
\end{figure*}

We repeated the above analysis on the full range of FLAMINGO 
L1 runs with hydrodynamics at $z=0$. For the runs that vary the 
cosmological model (including the neutrino component) we find, 
reassuringly, almost no change in the LLR parameters 
for the range of halo masses. For runs with varying 
resolution there are some differences at lower mass but the 
fiducial resolution (L1\_m9) is reasonably well converged 
(see Appendix~\ref{sec:res_study}). The runs with lower stellar masses 
(M*-1$\sigma$ and M*-1$\sigma$\_fgas-4$\sigma$) produce very similar 
results to their L1\_m9 and fgas-4$\sigma$ counterparts. This is 
because the hot gas dominates the baryon budget in group and 
cluster-sized haloes. 

The most significant systematic differences in the LLR parameters 
are found when the hot gas fractions are varied (relative 
to observational uncertainties), mainly driven by the strength of 
AGN feedback events (more energetic events, as a result of 
higher heating temperature, lead to lower gas fractions). 
Fig.~\ref{fig:tmrel_agn_llr} shows the LLR parameters for these 
runs at $z=0$ with fgas-8$\sigma$ (turquoise) using the 
strongest feedback and fgas+2$\sigma$ (purple) the weakest. 
As the gas fractions are lowered (through increasing the feedback 
strength), the gas, unsurprisingly, becomes hotter at fixed mass, 
with the SZ $y$-weighted temperatures still the highest of the 
three. 

We also see the mass scale where the slope is at a minimum 
increase when the gas fractions are lowered. A similar result was seen 
in the cosmo-OWLS simulations with varying AGN heating temperatures, 
studied by \cite{LeBrun2014}, and is due to the 
effect of the AGN feedback on the entropy of the gas.
At the lowest masses ($M_{\rm 500c} \sim 10^{13}\,{\rm M}_{\odot}$), 
the average mass-weighted temperatures are similar across 
the 5 gas fraction models (as can be seen from the normalisation, 
where the models also have $e^{\pi}>T_{\rm 500c}$, i.e. the intragroup 
gas is hotter than the gravitational temperature). 
Here, the feedback is effective at heating and ejecting gas from the halo 
since the AGN heating temperature, $\Delta T_{\rm AGN} \gg T_{\rm 500c}$. 
As halo mass increases towards cluster scales (and 
$T_{\rm 500c} \rightarrow \Delta T_{\rm AGN}$), the impact of the feedback 
on the gas temperature reduces, correspondingly leading to a decrease in the 
temperature ratio $e^{\pi}/T_{\rm 500c}$ and the slope, $\alpha$, also 
decreases. Eventually, the slope starts increasing again as the 
temperature becomes more and more dominated by gravitational heating 
at the largest masses. This transition occurs at a larger halo mass for a 
run with lower gas fractions due to its larger $\Delta T_{\rm AGN}$ value.

Models with lower gas fractions/stronger feedback also produce 
more scatter, especially in groups, and this also reaches a 
minimum value at a larger halo mass.
The scatter is particularly large for the strongest feedback 
case (fgas-8$\sigma$) due to groups with very high temperatures.
However, the scatter is much less model dependent 
for the most massive clusters where gravitational heating dominates.

Another set of runs with varying baryonic physics are those that 
use jet (i.e. directed kinetic) feedback rather than thermal 
feedback for the AGN. These models (Jet and Jet\_fgas-4$\sigma$) 
produce qualitatively similar results to their respective thermal 
feedback models (L1\_m9 and fgas-4$\sigma$) with the jet models 
producing slightly (up to around 10 per cent or so) 
lower temperatures and scatter on group scales. This is probably due 
to incomplete thermalization of feedback energy, possibly linked to  
the lower mass objects being less well resolved. 

\section{Radial Profiles and Hydrostatic Masses}
\label{sec:profmass}

Results in the previous section showed that the global SZ $y$-weighted 
temperatures are higher than both mass-weighted and 
X-ray spectroscopic-like temperatures at fixed halo mass, and vary 
with redshift at a rate that is closest to the self-similar 
expectation. These results are in agreement with previous work using 
smaller samples \citep[e.g.][]{Lee2022}. We also found the temperatures 
to be sensitive to variations in cluster gas fractions (primarily 
driven by AGN feedback strength) but insensitive to variations in 
galaxy stellar masses and the underlying cosmological/neutrino model.

We now investigate whether the use of relativistic SZ 
temperatures can reduce the hydrostatic mass bias 
in massive clusters; for this we require the $y$-weighted 
radial temperature profiles. For SZ-based data where gas thermal 
electron pressure profiles can also be extracted (from Compton-$y$ profiles), 
it is appropriate to express the hydrostatic mass as a function of 
temperature and pressure using the 
following version of the hydrostatic equilibrium equation
\begin{equation}
M(<r) = - {k_{\rm B} \over G \mu m_{p}} \, 
\left[ r \, T \, {{\rm d}\ln P \over {\rm d} \ln r} \right],
\label{eqn:hse}
\end{equation}
where the term in square brackets is a function of the radius $r$ 
and $P$ can be either the total or the electron thermal pressure since
the mass depends on the relative differential ${\rm d}\ln P = {\rm d}P/P$. 
We assume $\mu=0.59$ here, typical of observational analyses 
\citep[e.g.][]{Eckert2019}.
Thus, the estimated hydrostatic mass, $M_{\rm 500c,hse}$, requires 
both the local temperature and local pressure gradient 
at $R_{\rm 500c,hse}$; we will outline the procedure below and refer 
the reader to \cite{Braspenning2023} for a more detailed study of the 
ICM profiles in comparison with X-ray data.
Note that we only use 3D profiles here although, in practice, the observed 
profiles are projected along the line-of-sight. We choose this 
approach as it allows us to focus on the effect of varying the 
temperature weighting on the hydrostatic mass estimate
(and other underlying gas properties). 
Observational predictions will also require projection 
effects to be taken into account (e.g. from simulating an 
SZ lightcone) as well as other complications such as non-SZ 
sources (including the CMB) and noise. Such predictions are better 
focused on specific instruments (taking into account the 
available frequency channels, beam size etc.) and 
are beyond the scope of the current work. However, for now, 
we note that \cite{Lee2020} showed that their results for 
projected (cylindrical) temperature profiles produced similar 
differences between the different temperature weightings for 
the high-mass clusters relevant to this study.

For the main analysis, we use the L2p8\_m9 run and only select 
clusters at $z=0$ with 
$M_{\rm 500c}>7.5 \times 10^{14}\,{\rm M}_{\odot}$ 
(1253 objects in total; 461 clusters have 
$M_{\rm 500c}>10^{15}\,{\rm M}_{\odot}$). This is because the 
most massive clusters have the hottest gas (on average) and thus 
produce the largest relativistic SZ signal. 
Our mass limit also ensures we 
still have a reasonable number of objects in the smaller 
L1 boxes that we use for comparing 
results with varying models (48 clusters for L1\_m9).
Our choice of redshift maximises the number of high-mass 
systems in our box. However, in practice, observations of massive 
clusters are likely to be at intermediate redshift. For example, 
RX J1347.5-1145, as studied recently by \cite{Butler2022}, is at 
$z=0.45$, while the CHEX-MATE Tier-2 sample \citep{chexmate2021} is 
at $0.2<z<0.6$. Therefore, we also compare our main results 
(hydrostatic bias parameters at $z=0$) to those at $z=0.5$; 
the latter sample being an order of magnitude smaller with 181 
objects.

\subsection{Pressure profiles}
\label{subsec:presprof}

\begin{figure}
\centering
\includegraphics[scale=0.45]{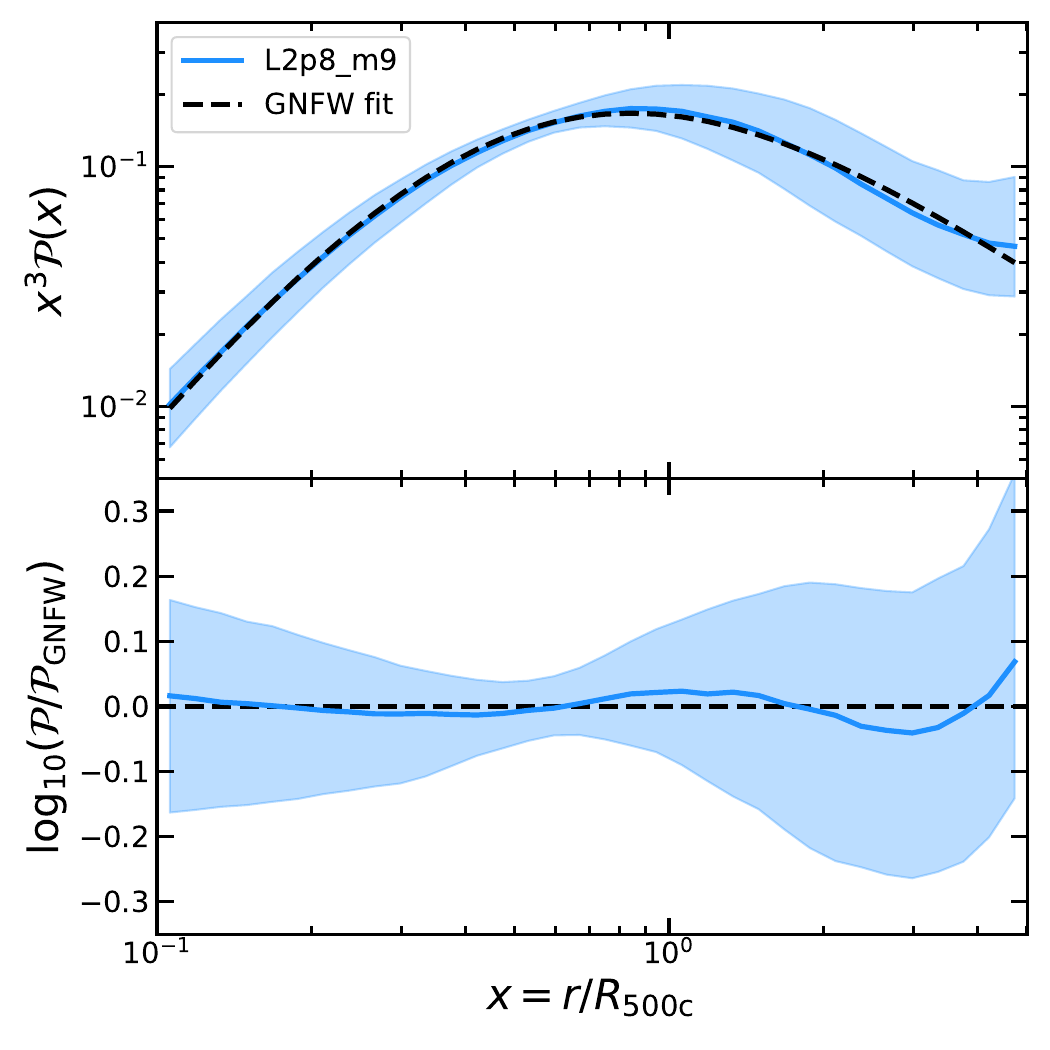}
\caption{
Top: scaled pressure profiles, plotted as $x^3\mathcal{P}(x)$, 
for the massive 
($M_{\rm 500c}>7.5\times 10^{14}\,{\rm M}_{\odot}$) clusters 
in L2p8\_m9 at $z=0$. The solid curve is the median profile 
while the shaded region shows the scatter (16th-84th percentiles). 
The dashed curve is the best-fitting 
generalized NFW (GNFW) model to the median profile. 
Bottom: results are shown relative to the best fit 
GNFW model.
}
\label{fig:prof3D_Pe}
\end{figure}

We show the (volume-weighted) thermal electron pressure profiles for the L2p8\_m9 
massive cluster sample in Fig.~\ref{fig:prof3D_Pe}.
The median scaled pressure profile, 
$\mathcal{P} = P/P_{\rm 500c}$, 
is plotted as a function of the dimensionless radius
$x=r/R_{\rm 500c}$ (solid curve), along with the 
16th to 84th percentiles (shaded region). We scale the 
y-axis by $x^{3}$ to highlight the relative contribution to the 
ICM thermal energy (or $Y$) from each radial bin as well as to 
highlight differences between the data and model. We do not remove any 
substructures when calculating the pressure profile as this would be a 
difficult thing to do in practice with SZ data, given the relatively 
low angular resolution of the observations.

The median profile is fitted with the generalized NFW 
(GNFW; \citealt{Nagai2007b}) model
\begin{equation}
    \mathcal{P}_{\rm GNFW}(x) = {P_{0} \over
    \left( c_{500}x \right)^{\gamma}
    \left[ 1 + (c_{500}x)^{\alpha}\right]^{(\beta-\gamma)/\alpha}
    },
\end{equation}
over the radial range $0.1<x<5$ (as shown). This range excludes the 
inner region so does not constrain the inner slope parameter, 
$\gamma$, well. Thus, following \cite{Barnes2017}, we fix its value to 
$\gamma=0.31$ (as found by \citealt{Arnaud2010}) and fit the other 
4 parameters 
$\{ P_{0}=6.05 \pm 0.32,
c_{500}=1.77 \pm 0.09,
\alpha=1.48 \pm 0.08,
\beta=4.55 \pm 0.12 \}$ 
where the given values are for the best-fitting model (shown as the 
dashed curve). This produces a reasonably good 
fit to the median profile, in line with previous simulation studies 
\citep[e.g.][]{Nagai2007b,Kay2012,Barnes2017,Gupta2017,Planelles2017}; 
with the largest deviations occurring beyond $R_{\rm 500c}$.

\begin{figure}
\centering
\includegraphics[scale=0.45]{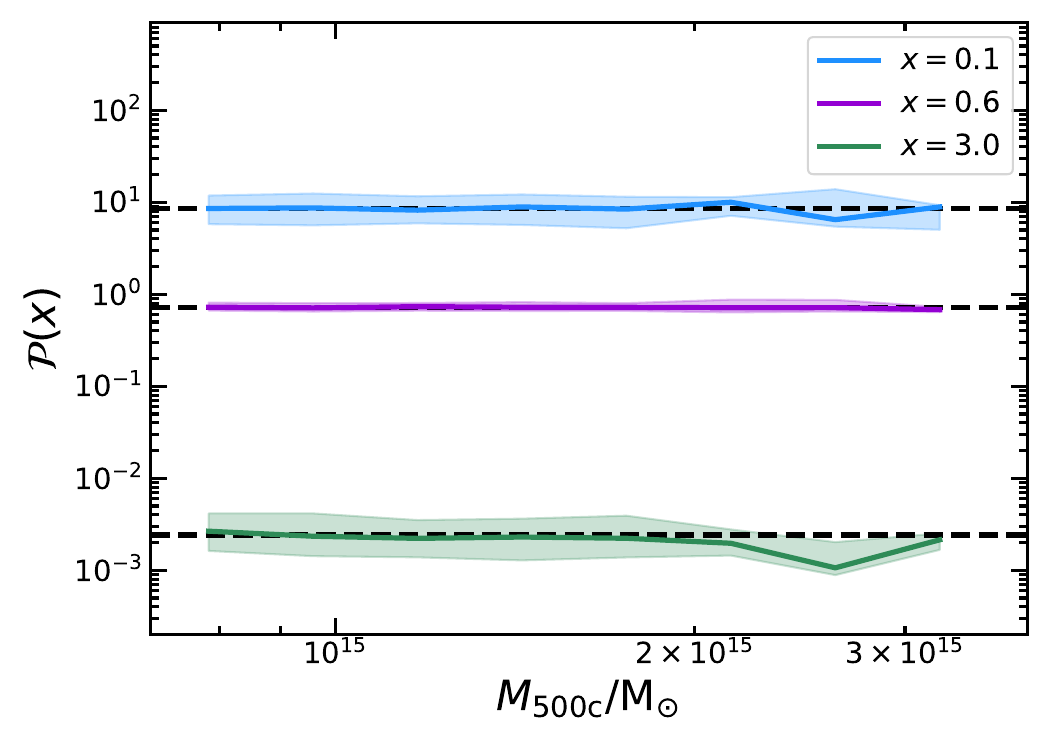}
\caption{
Scaled pressure values at 3 different radii versus halo mass. 
Each solid line corresponds to the median pressure 
while the shaded region shows the scatter (16th-84th 
percentiles). The median pressure of the sample  
is shown by the horizontal dashed line in each case. 
The radii were chosen to show the pressure values 
in the core ($x=0.1$); 
at intermediate radius where the scatter is minimal ($x=0.6$) 
and in the outskirts where the scatter is maximal ($x=3$).
These results show that the scatter is not driven by an additional 
dependence of the pressure on cluster mass.
}
\label{fig:prof3D_Pe_M500c}
\end{figure}

Looking at the scatter, we find it is minimal at 
$x \approx 0.6$ and maximal at $x \approx 3$. 
To check whether this scatter is 
due to an additional mass dependence (over and above the self-similar 
scaling), we show in Fig.~\ref{fig:prof3D_Pe_M500c} the scaled pressure 
values for individual clusters versus their mass at 3 different radii: 
$x=0.1$ (the core); 
$x=0.6$ (intermediate radius, minimal scatter) and 
$x=3$ (outskirts, maximal scatter). It is clear the scaled pressure 
has weak or no dependence on halo mass, with the median value close 
to the sample median (dashed line) in all cases except for a deviation 
in the most massive clusters 
($M_{\rm 500c}>2\times 10^{15}\,{\rm M}_{\odot}$) at $x=0.1$ and $x=3$.
The Spearman correlation coefficients are $r_{s}=(0.01,-0.007,-0.2)$ 
for $x=(0.1,0.6,3)$, respectively. The larger scatter at large radius is 
likely to be associated, at least in part, with merger shocks (see below).

\begin{figure}
\centering
\includegraphics[scale=0.45]{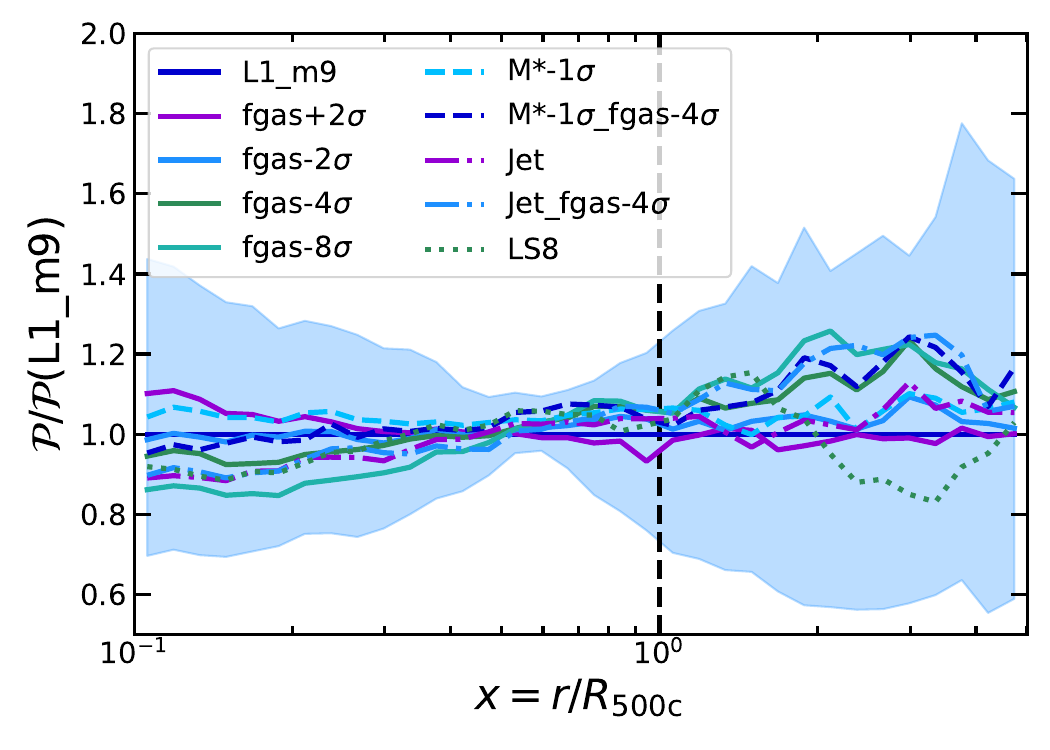}
\caption{A comparison of the median pressure 
profiles for clusters with $M_{\rm 500c}>7.5 \times 10^{14}\,{\rm M}_{\odot}$
in the different L1 models (see text for details), relative to 
the L1\_m9 case. The shaded region illustrates 16th-84th percentiles for 
L1\_m9, showing that the level of cluster-to-cluster scatter is larger 
than the variations between models.}
\label{fig:pprof3d_comp}
\end{figure}

In Fig.~\ref{fig:pprof3d_comp}, we compare the electron pressure profiles for 
clusters in a selection of runs with varying baryonic physics models, and the 
alternative LS8 cosmological model. In all cases, we see only 
relatively small differences ($<30$ per cent) in the pressure between 
models at each radius. In runs with lower hot gas fractions, driven 
by stronger AGN feedback, the pressure is lower in the inner region 
but higher in the outskirts, as may be expected from the ejection of 
more material to larger radii. Clusters within the LS8 run have lower 
pressure in the outskirts but this may be a statistical effect 
due to there being fewer objects above the mass limit in this run 
(28 clusters, compared with 48 in L1\_m9).

\subsection{Temperature profiles}

\begin{figure}
\centering
\includegraphics[scale=0.45]{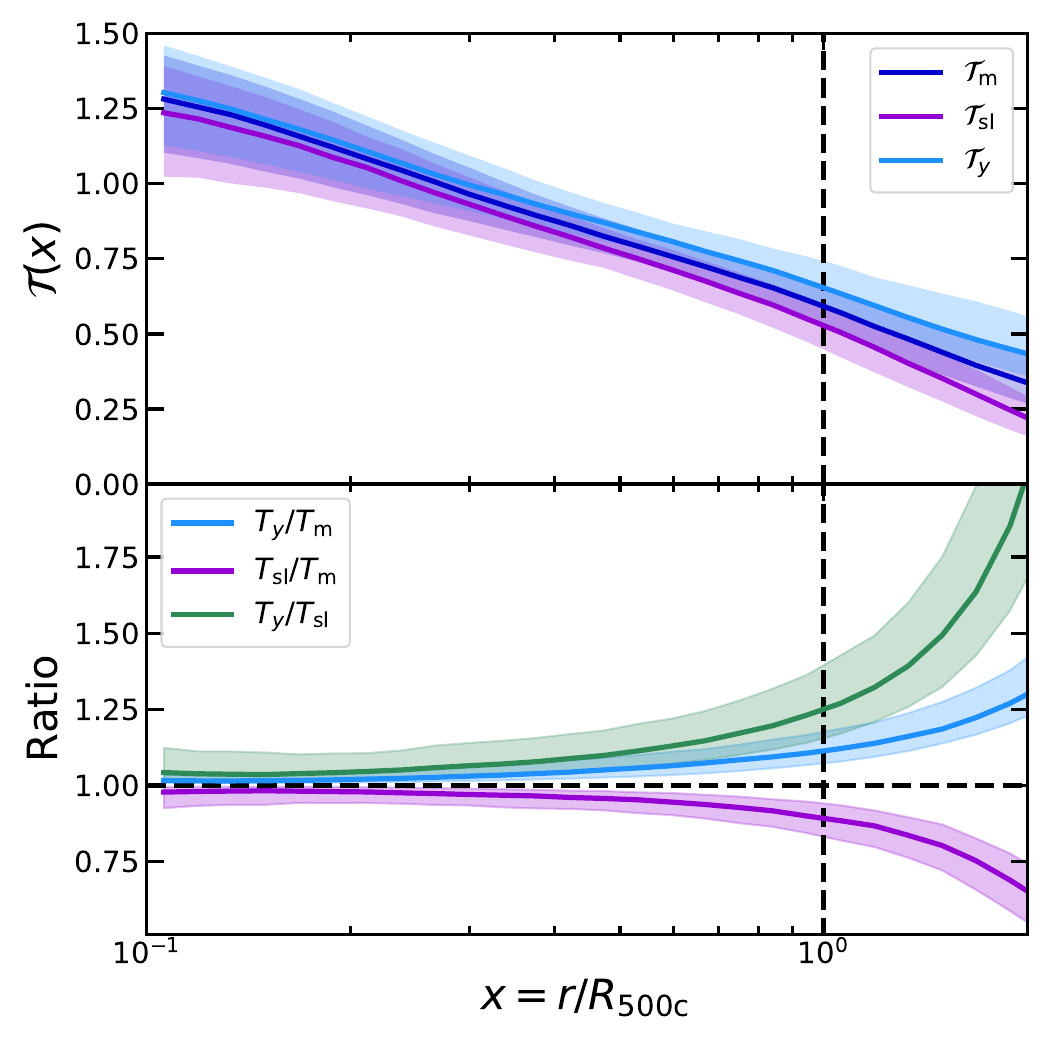}
\caption{Scaled temperature profiles for the massive clusters 
in L2p8\_m9 at $z=0$, in the radial range $0.1<x<2$. The top panel 
shows results for the 3 different temperature weightings  
(mass weighted, spectroscopic-like weighted and $y$ weighted) while the 
bottom panel shows temperature ratios between two weightings. 
The solid curves are the median profiles 
while the shaded regions span the 16th-84th percentiles. The dashed 
vertical line highlights $x=1$ ($r=R_{\rm 500c}$).}
\label{fig:prof3D_T}
\end{figure}

We next compare the 3D radial $y$-weighted temperature profiles 
to the mass-weighted and spectroscopic-like cases. 
Fig.~\ref{fig:prof3D_T} shows the median temperature profiles 
and the 16th-84th percentile regions from 
L2p8\_m9. We restrict the radial range to $0.1<x<2$ as we 
are most interested in the temperatures around $x=1$ for 
calculating the hydrostatic masses. Furthermore, it will be much 
more observationally challenging to measure temperatures at 
$x \gg 1$ where the SZ signal is lower, the gas is cooler and the 
physics is more complex (e.g. the electron and ion temperatures 
may no longer be equal, see e.g. \citealt{Fox1997}).

\begin{figure}
\centering
\includegraphics[scale=0.45]{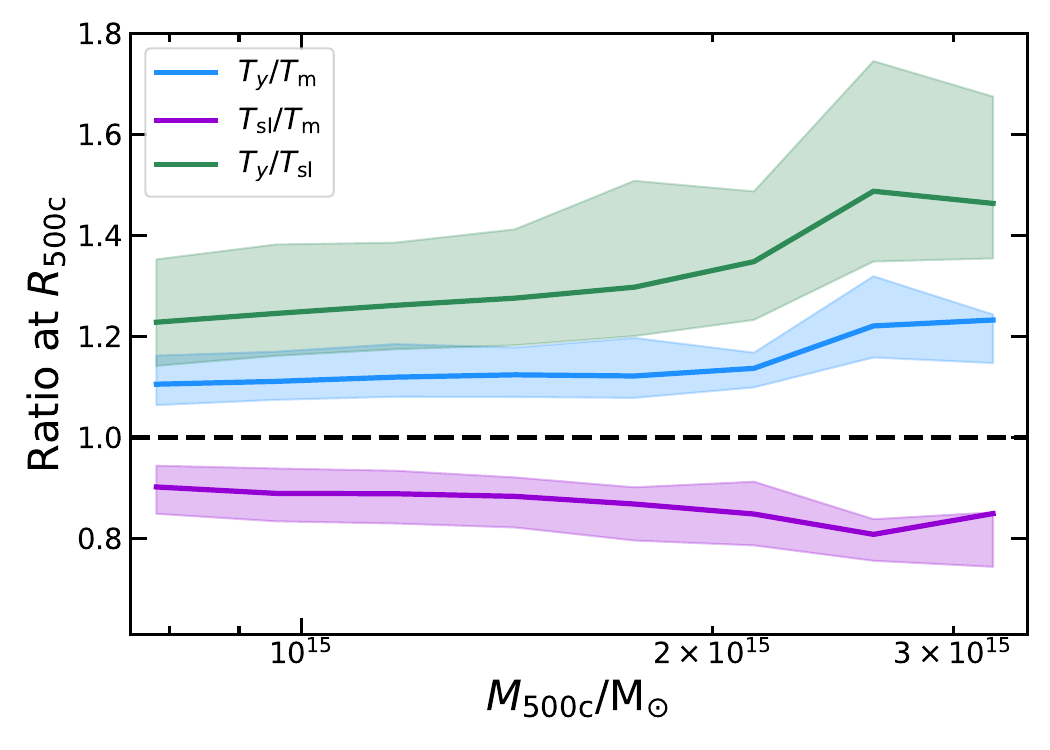}
\caption{Ratio of temperatures measured at 
$R_{\rm 500c}$ versus halo mass. Solid curves show the median ratios 
and shaded regions the 16th-18th percentiles for each mass 
bin. The dashed horizontal line illustrates a ratio of unity.}
\label{fig:tprof3d_ratio}
\end{figure}

In line with our $T-M$ scaling results and with 
previous work \citep{Kay2008,Lee2020,Lee2022}, 
we find that the $y$-weighted temperature, $T_{y}$, is 
larger than the mass-weighted temperature, $T_{\rm m}$, and 
the spectroscopic-like temperature, $T_{\rm sl}$, at all 
radii considered. Within the main cluster region 
these differences are relatively modest, with 
$T_{y}$ and $T_{\rm sl}$ being within $10-15$ per cent of 
$T_{\rm m}$ at $x=1$ ($r=R_{\rm 500c}$), 
but they increase significantly in the cluster outskirts ($1<x<2$). 
As can be seen in the lower panel, the $y$-weighted temperature is 
around twice the spectroscopic-like temperature and 25 per cent larger 
than the mass-weighted temperature at $x=2$.
We will explore the outskirts further below, but for now, we note that the 
offsets around $R_{\rm 500c}$ will affect the mass estimates, given that 
$M \propto T$ in equation~\ref{eqn:hse}. 
Fig.~\ref{fig:tprof3d_ratio} compares these ratios at $R_{\rm 500c}$,  
showing the median values and the scatter as a function of halo 
mass. While most of the objects are at the lower-mass end, the trend 
is for the highest-mass clusters to have $T_y/T_{\rm m}$ ratios 
that are around 10 per cent higher, and $T_{\rm sl}/T_{\rm m}$ ratios 
that are around 10 per cent lower, than the lowest-mass objects. 
Consequently, the median $T_{y}/T_{\rm sl}$ ratio (SZ/X-ray temperature) 
varies by around 20 per cent or so, over the same mass range. Spearman 
correlation coefficients are $r_{s}=(0.2,-0.2,0.2)$ for the 
$(T_{y}/T_{\rm m}, T_{\rm sl}/T_{\rm m}, T_{y}/T_{\rm sl})$ 
ratios with $M_{\rm 500c}$, respectively.

\begin{figure}
\centering
\includegraphics[scale=0.45]{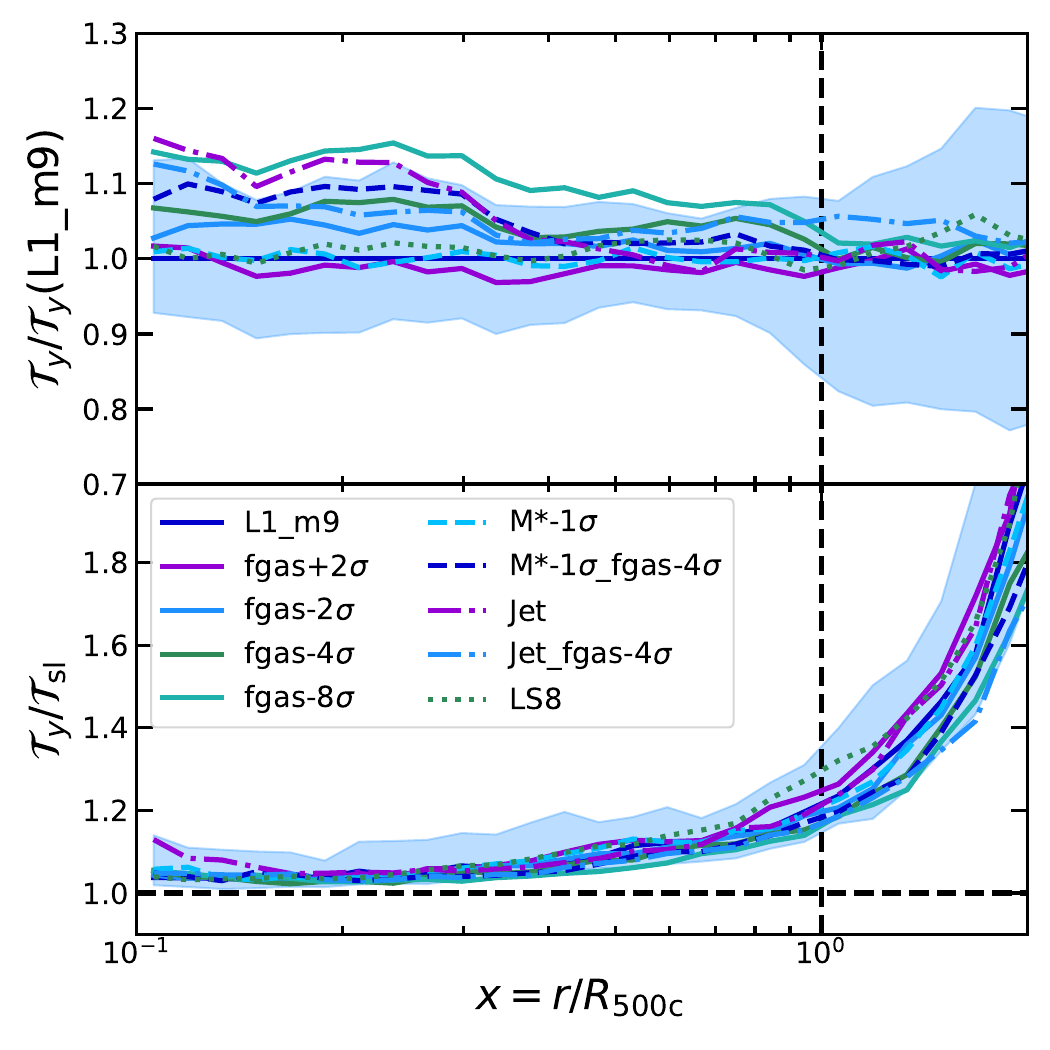}
\caption{A comparison of the median temperature 
profiles between different L1 models. Top: ratio of 
$y$-weighted temperature profiles to the fiducial (L1\_m9) case. 
Bottom: ratio of $y$-weighted and 
spectroscopic-like profiles. The shaded region illustrates 
16th-84th percentiles for L1\_m9, showing the level of cluster-to-cluster 
scatter.}
\label{fig:tprof3d_comp}
\end{figure}

Fig.~\ref{fig:tprof3d_comp} shows results from comparing the temperature 
profiles for the different L1 models. In the top panel, we can see that 
relative differences in $T_{y}$ between the models and the fiducial 
case mainly occur within $R_{\rm 500c}$. As the hot gas fraction 
decreases (from increasing the thermal AGN feedback strength) the 
temperature within the cluster increases but the effect is mild 
(within 20 per cent in the extreme, fgas-8$\sigma$ case). Interestingly, 
the jet model, using directed kinetic AGN feedback, produces a similar  
increase that almost reaches 20 per cent at $0.1R_{\rm 500c}$ 
(see also \citealt{Braspenning2023} who compared mass-weighted profiles). 
Beyond $R_{\rm 500c}$, $T_{y}$ is less sensitive to the feedback 
variations with differences less than 5 per cent. Models with 
varying stellar masses and cosmology/neutrinos show no discernible 
systematic differences in $T_{y}$ profiles (for the latter, only 
the LS8 case is shown here).

We also show the ratio of median $T_{y}$ and $T_{\rm sl}$ profiles 
for individual models in the bottom panel. 
These results show that the ratio is fairly insensitive to the feedback 
variations out to $2R_{\rm 500c}$. We can see that, in the cluster 
outskirts ($r>R_{\rm 500c}$) where the ratio is larger, differences 
between the models are smaller than the cluster-to-cluster scatter 
for L1\_m9. This suggests the large offset between the two temperatures 
is a robust prediction from these simulations and so it would be 
interesting to test this with X-ray and (relativistic) SZ 
observations e.g. by stacking clusters, once it is possible to 
measure temperatures at these radii. However, it is not clear whether 
this prediction, based on simple 3D weighted temperatures, would 
accurately reflect the observed temperature ratio on these scales where the  
gas is intrinsically cooler and projection effects are likely to be 
significant. To test this, we would require more detailed modelling using 
mock observations, something that we leave to future work.

\subsection{Hydrostatic masses}

We now estimate individual cluster masses using 
equation~\ref{eqn:hse}. Since the individual profiles can be 
quite noisy (often due to substructure producing localised 
fluctuations), we fit model profiles to the 
pressure and temperature. For the pressure, we fit a 4-parameter 
GNFW model, fixing $\gamma=0.31$ as in Section~\ref{subsec:presprof}.
Here, we also restrict the radial range to $0.1<x<2$, to avoid the largest 
radii where deviations from a smooth profile are larger. For the 
temperature profile, it is common to model this using the function 
described in \cite{Vikhlinin2006}. However, as we only need the 
temperature around $R_{\rm 500c}$, such a model (with up to 
7 parameters) is over-complicated for our needs. Instead, we found 
a simpler, third-order (cubic) polynomial function to be sufficient 
when applied over the radial range $0.5<x<2$. These fits were 
made for each of the three temperature profiles ($T_{\rm m}$, 
$T_{\rm sl}$ and $T_{y}$). 

We perform these fits for each of the 1253 clusters 
with $M_{\rm 500c}>7.5\times 10^{14}\,{\rm M}_{\odot}$ in L2p8\_m9, 
using the pressure and temperature models to calculate the mass
profile for $0.5<x<2$ using equation~\ref{eqn:hse}. From the 
estimated mass, we then calculate the radius at which the 
mean internal density, 
$\left< \rho \right>=500\rho_{\rm cr}$. 
This radius is labelled $R_{\rm 500c,hse}$ and the estimated 
mass $M_{\rm 500c,hse}$. The hydrostatic mass bias at 
$R_{\rm 500c,hse}$, $b$, is then defined through
\begin{equation}
    1-b={M_{\rm 500c,hse} \over M_{\rm 500c}}. 
\label{eqn:biasdef}
\end{equation}

\begin{figure}
\centering
\includegraphics[scale=0.45]{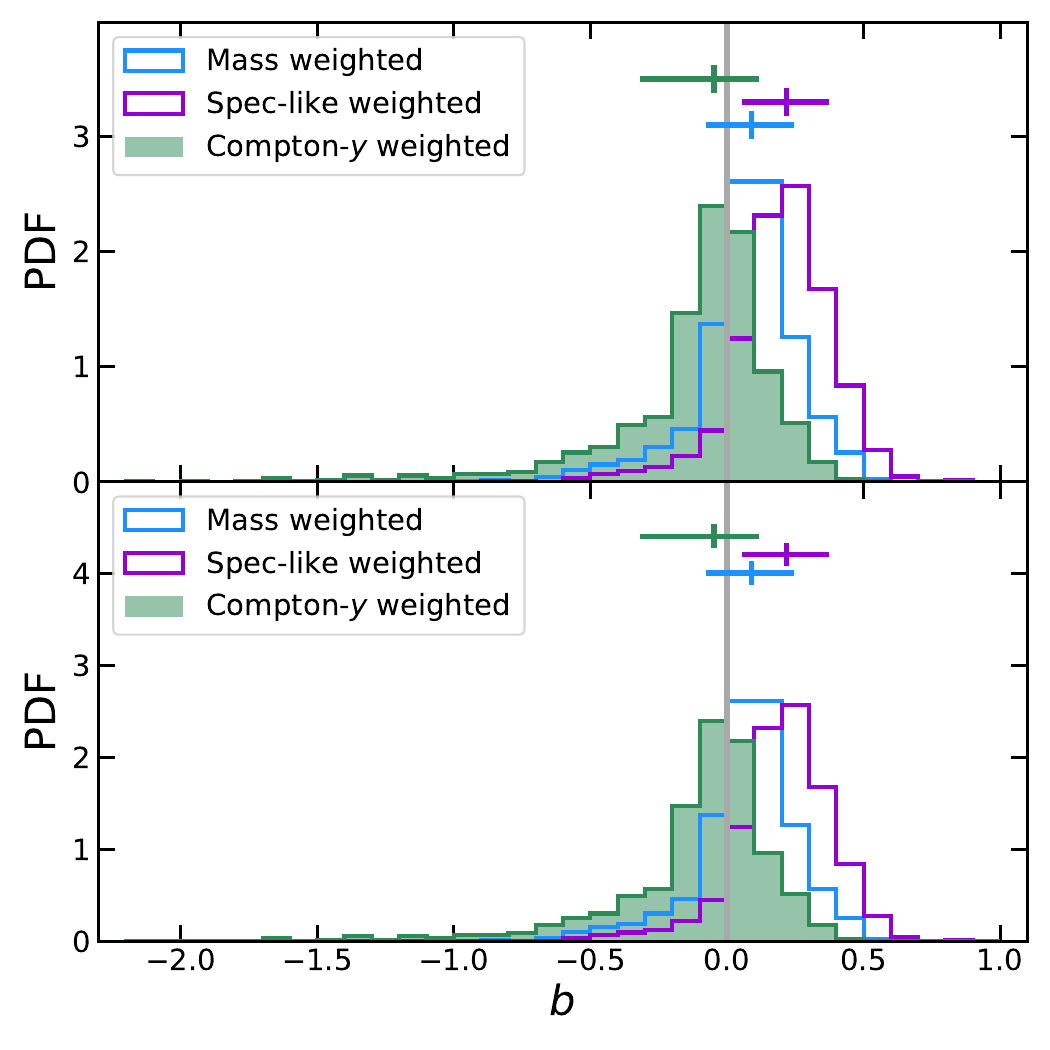}
\caption{
Top: hydrostatic mass bias ($b$; equation~\ref{eqn:biasdef}) 
distributions for massive clusters in the L2p8\_m9 run at $z=0$. 
The three histograms correspond to the different temperature 
profile weightings, as shown in the legend. The grey vertical 
line denotes $b=0$ (no bias) and the crosses illustrate the 
median and 16th-84th percentiles for each distribution. 
Bottom: as above but for the subset with the lowest 
pressure profile goodness-of-fit values, $\Delta(\mathcal{P})<0.02$. 
}
\label{fig:hse_bias}
\end{figure}

\begin{table}
	\centering
	\caption{Median hydrostatic bias values and their scatter 
 for the L2p8\_m9 massive ($M_{\rm 500c}>7\times 10^{14}\,{\rm M}_{\odot}$)
 cluster sample and the subsample with the best 
 GNFW pressure profile fits ($\Delta(\mathcal{P})<0.02$). 
 Results are given for both $z=0$ 
 (1253 clusters) and $z=0.5$ (181 clusters).
 Uncertainties are calculated from 
 bootstrap re-sampling 10,000 times.}
	\label{tab:bias_values}
	\begin{tabular}{lcc} 
		\hline
        Weighting & $\left< b \right>$ & $\sigma_{b}$ \\
		\hline
        {\bf All clusters} $z=0$: &&\\
        Mass & $\phantom{-}0.091 \pm 0.005$ & $0.152 \pm 0.006$\\
        Spec-like & $\phantom{-}0.217 \pm 0.005$ & $0.151 \pm 0.004$\\
        Compton-$y$ & $-0.048 \pm 0.005$ & $0.209 \pm 0.009$\\
        {\bf All clusters} $z=0.5$: &&\\      
        Mass & $\phantom{-}0.107 \pm 0.013$ & $0.181 \pm 0.014$\\
        Spec-like & $\phantom{-}0.281 \pm 0.012$ & $0.139 \pm 0.011$\\
        Compton-$y$ & $-0.160 \pm 0.028$ & $0.306 \pm 0.036$\\
		\hline
        {\bf Good-fit clusters} $z=0$: &&\\ 
        Mass & $\phantom{-}0.114 \pm 0.006$ & $0.112 \pm 0.006$\\
        Spec-like & $\phantom{-}0.226 \pm 0.007$ & $0.125 \pm 0.005$\\
        Compton-$y$ & $-0.009 \pm 0.005$ & $0.115 \pm 0.005$\\
        {\bf Good-fit clusters} $z=0.5$: &&\\ 
        Mass & $\phantom{-}0.156 \pm 0.013$ & $0.088 \pm 0.015$\\
        Spec-like & $\phantom{-}0.340 \pm 0.022$ & $0.110 \pm 0.013$\\
        Compton-$y$ & $-0.036 \pm 0.017$ & $0.100 \pm 0.018$\\
        \hline
	\end{tabular}
\end{table}

\begin{figure*}
\centering
\includegraphics[scale=0.45]{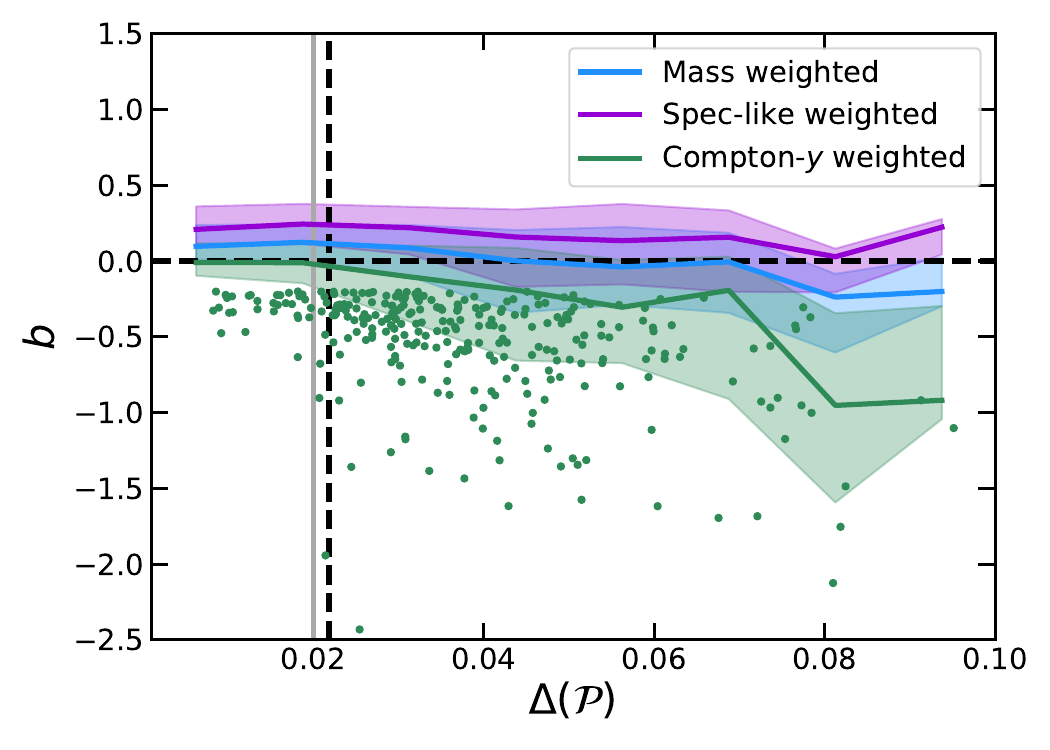}
\includegraphics[scale=0.45]{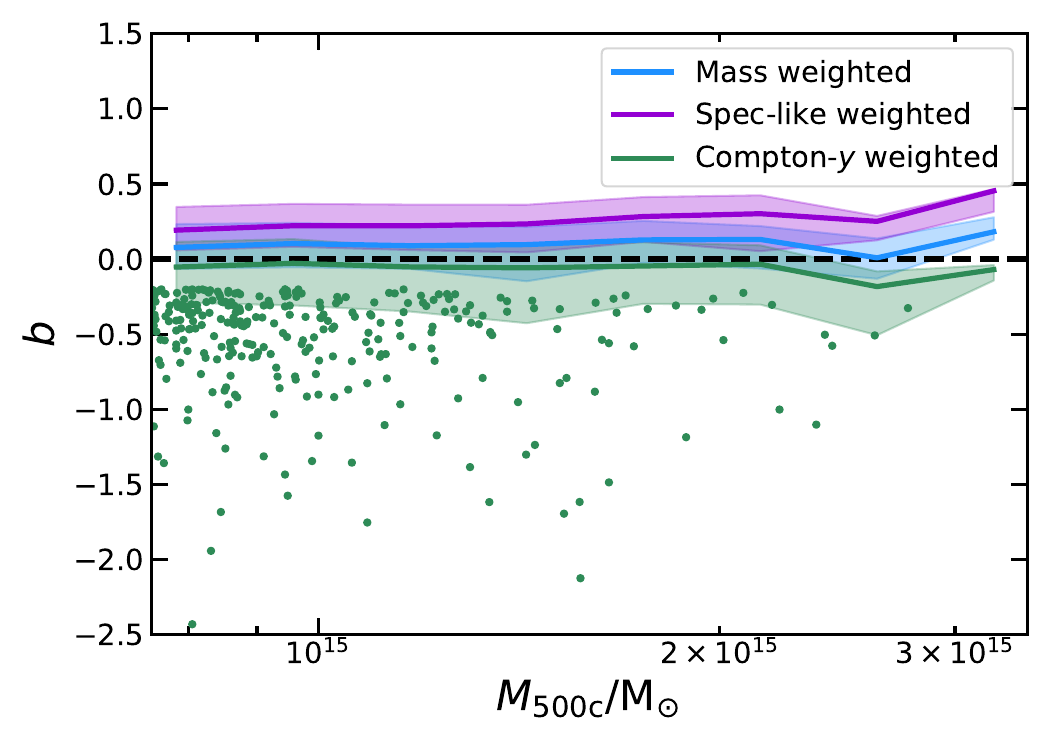}
\caption{
Hydrostatic mass bias, $b$, versus pressure profile goodness-of-fit 
statistic, $\Delta(\mathcal{P})$ (left), and halo mass, 
$M_{\rm 500c}$ (right), for the massive cluster sample at $z=0$. 
The three temperature cases are shown in different colours, with 
the solid lines showing the median bias and the shaded regions 
the 16th-84th percentiles. The green dots show results for 
individual clusters with $b<-0.2$ for the $y$-weighted case. 
The dashed vertical line is the median value of $\Delta(\mathcal{P})$ 
while the solid grey vertical line corresponds to 
$\Delta(\mathcal{P})=0.02$, the upper limit used to define the subset with 
good pressure profile fits.
}
\label{fig:bias_gof}
\end{figure*}

\begin{figure*}
\centering
\includegraphics[scale=0.55]{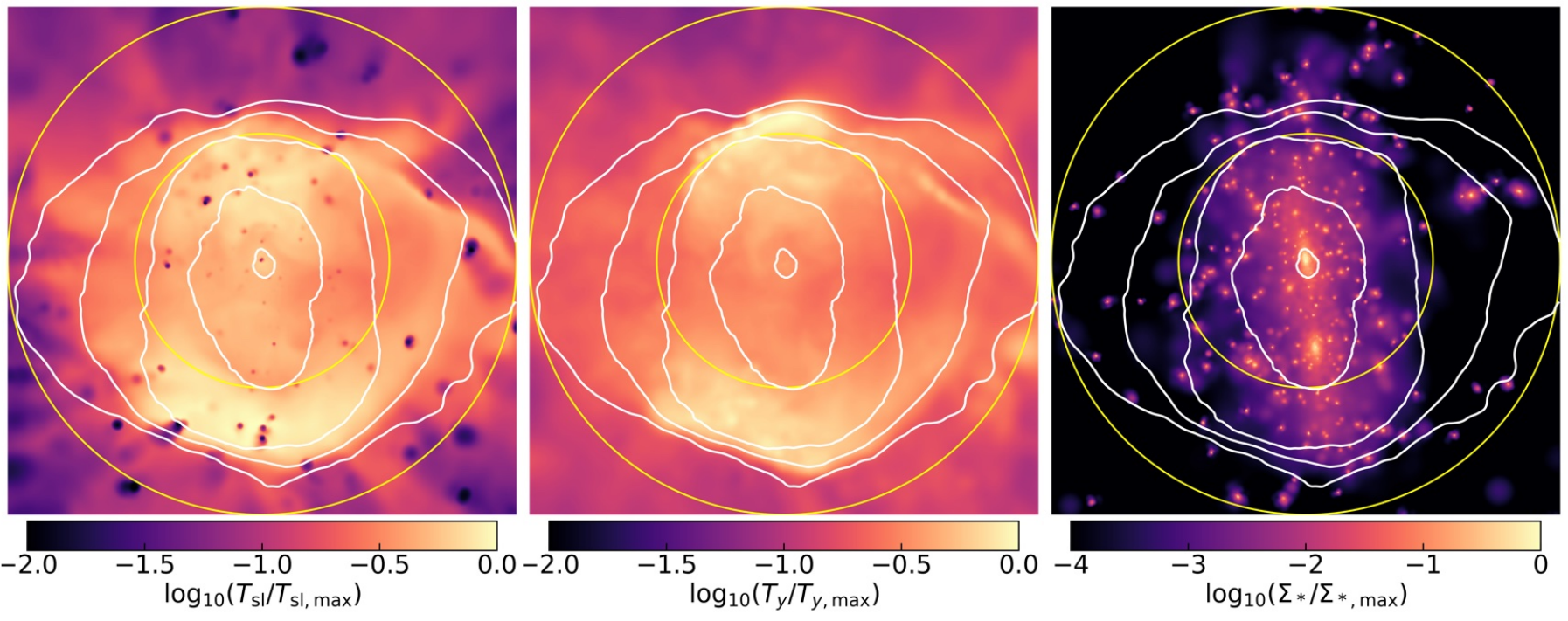}
\caption{
Spectroscopic-like temperature (left), Compton-$y$ weighted temperature 
(middle) and stellar mass density (right) maps for the cluster with 
a large negative mass bias using $y$-weighted temperatures 
($b=-2.1$). The maps show the projection of a cubic region with side 
length $4R_{\rm 500c}$, centred on the cluster. Pixel values are 
normalised to the maximum and are shown on a logarithmic scale 
to improve contrast. The white contours represent equal Compton-$y$ 
values, $\log_{10} (y/y_{\rm max})=[-2,-1.5,-1,-0.5,-0.1]$ 
whereas the yellow circles indicate $[1,2]R_{\rm 500c}$.
}
\label{fig:temp_maps}
\end{figure*}

Fig.~\ref{fig:hse_bias} (top panel) shows the distribution of $b$ values 
for the 3 different temperatures (the vertical grey 
line corresponds to $b=0$). 
The crosses above the histograms show the median and 
16th-84th percentiles (see also Table~\ref{tab:bias_values}). 
As in previous work (e.g. \citealt{Biffi2016,Henson2017,Pearce2020}), 
using the mass-weighted temperature profile 
results in a slightly smaller mass on average 
(the median bias is $\left< b \right> = 0.091\pm 0.005$) 
but the bias more than doubles when using the spectroscopic-like temperature 
($\left< b \right> = 0.217 \pm 0.005$) as a result of the latter being 
biased towards denser, cooler gas. Both distributions have 
similar scatter, with $\sigma_{b} \approx 0.15$ where 
$\sigma_{b}$ is defined to be half the difference between the 
16th and 84th percentile $b$ values. On the other hand, 
using the $y$-weighted temperature results in a lower bias with  
$\left< b \right> = -0.048 \pm 0.005$ 
but larger scatter, with $\sigma_{b} = 0.21 \pm 0.01$. 
This systematic shift in $\left< b \right>$ is expected given 
that $T_{y}$ is larger than the other two measures. The 
larger scatter is mainly due to the tail of low $b$ values 
(the lowest value, an extreme outlier, has $b=-2.4$). While this 
tail is present in all 3 distributions, 
it is most prominent in the $y$-weighted case, suggesting its 
origin must be related to the thermal pressure of the gas 
around $R_{\rm 500c}$.

To investigate this tail, we calculate a goodness-of-fit statistic for the 
GNFW fit to the pressure profile for each cluster
\begin{equation}
    \Delta(\mathcal{P}) = 
    \sqrt{
    \frac{1}{N_{\rm bins}}
    \sum_{i=1}^{N_{\rm bins}} 
    \left[
    \log_{10} 
    \left( \mathcal{P}/\mathcal{P}_{\rm GNFW} \right)
    \right]^{2}
    },
\end{equation}
where $N_{\rm bins}=26$ is the number of radial bins in the range 
$0.1<x<2$ used for the GNFW fit. We plot $b$ versus $\Delta(\mathcal{P})$ 
for the massive cluster sample in the left panel of 
Fig.~\ref{fig:bias_gof}. This shows that clusters with higher $\Delta(\mathcal{P})$ values (poorer pressure profile fits) tend to have 
larger $b$ scatter. This is particularly significant in the $T_{y}$ case 
where we additionally show the individual clusters with the most 
negative bias values ($b<-0.2$) as dots. We also note an overall trend 
in the median $b$ decreasing with increasing $\Delta(\mathcal{P})$; 
the Spearman coefficients are $r_{s}=(-0.2,-0.1,-0.3)$ when using 
$T_{\rm m}, T_{\rm sl}, T_{y}$, respectively. This confirms our expectation  
that the tail of negative $b$ values primarily contains objects with poor 
pressure profile fits. 

We also show the $b$ values as a function of halo 
mass in the right panel of Fig.~\ref{fig:bias_gof}. There is a mild 
positive correlation in the $T_{\rm sl}$ case ($r_{s}=0.11$), 
as seen in previous work 
\cite[e.g.][]{Barnes2021}, but this is considerably weaker 
in the other two cases ($r_{s}=0.03$ for $T_{\rm m}$ and $r_{s}=-0.03$ for 
$T_{y}$). The scatter shows no obvious trend with halo mass; clusters 
with $b<-0.2$ are found across our (limited) halo mass range but are 
more common at lower mass where there are more objects in total.

An example of a cluster with a poor pressure profile fit 
($\Delta(\mathcal{P}) \approx 0.08$) and large, negative 
$y$-weighted bias ($b=-2.1$) is shown in Fig.~\ref{fig:temp_maps}. 
Each panel shows, 
from left to right, $T_{\rm sl}$, $T_{y}$ and stellar mass density 
maps, projected down one axis of a cube, centred on the cluster, with 
side length $4R_{\rm 500c}$. White contours illustrate Compton-$y$ values 
and the yellow circles $[1,2]R_{\rm 500c}$.  In this case (and typically 
for clusters with large $\Delta(\mathcal{P})$ values) there is clear 
evidence of dynamical activity: the stellar density map shows a double 
peak along the vertical direction, while the gas shows regions of 
large pressure gradients and high temperature perpendicular to 
this direction, associated with merger shocks. Note that these 
shocks occur between $1-2R_{\rm 500c}$ (between the two yellow circles) 
and hence affect the pressure profile fit.

\begin{figure}
\centering
\includegraphics[scale=0.45]{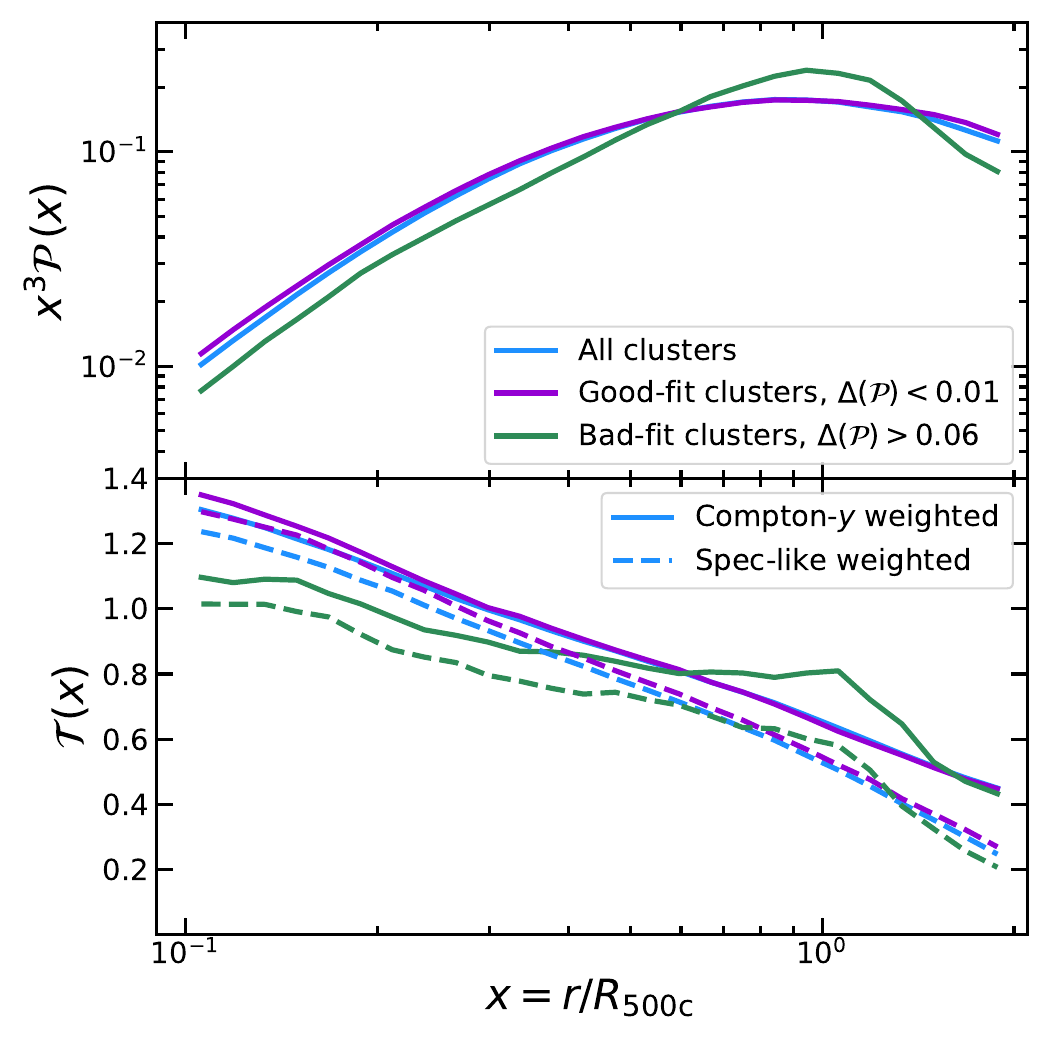}
\caption{
Median pressure (top) and temperature (bottom) profiles for 
all massive clusters in L2p8\_m9 (blue), compared to the subsamples with 
good (purple) and bad (green) fits to their GNFW model pressure profiles. 
}
\label{fig:profiles_goodbad}
\end{figure}

More quantitatively, we show median pressure and temperature profiles 
in Fig.~\ref{fig:profiles_goodbad} for all clusters (blue), clusters 
with good pressure profile fits ($\Delta(\mathcal{P})<0.01$; purple) 
and clusters with bad fits ($\Delta(\mathcal{P})>0.06$; green). 
In the bottom panel, solid curves are for the $y$-weighted temperature and 
dashed curves for the spectroscopic-like temperature.
The median pressure profile for the bad-fitting clusters is 
significantly different from the other two at nearly all radii.
The pressure is lower at $r<0.5R_{\rm 500c}$ and up to 30 
per cent higher at $r \approx R_{\rm 500c}$. The $y$-weighted 
temperature profile shows similar behaviour whereas 
the effect on the spectroscopic-like temperature profile is smaller
in the outskirts (around 10 per cent enhancement at $R_{\rm 500c}$). 
Consequently, it is these local increases in pressure and 
$y$-weighted temperature, associated with the merger shocks seen 
in Fig.~\ref{fig:temp_maps}, that lead to the tail of low (negative) 
$b$ values.

Removing clusters with the largest pressure deviations (relative 
to the GNFW model) would therefore be a simple, if not optimal, way 
to reduce the scatter in the mass bias. We demonstrate this by 
taking the subset of 563 clusters with $\Delta(\mathcal{P})<0.02$, close 
to the sample median (see Fig.~\ref{fig:bias_gof}). The resulting 
$b$ distributions for this subsample are shown in the lower panel in 
Fig.~\ref{fig:hse_bias}, with $\left<b\right>$ and $\sigma_{b}$ values 
listed in Table~\ref{tab:bias_values}. 
As expected, removing these clusters has the largest impact on 
the $y$-weighted case, where the median bias reduces to  
$\left< b \right> = -0.009 \pm 0.005$ and the scatter 
reduces by a factor of two, to $\sigma_{b} = 0.115 \pm 0.005$. 
For the mass-weighted case we find 
$\left< b \right> = 0.114 \pm 0.006$ and $\sigma_{b}=0.112 \pm 0.006$, 
while for the spectroscopic-like case 
$\left< b \right> = 0.226 \pm 0.007$ and $\sigma_{b} = 0.125 \pm 0.005$.
In these latter two cases, the median bias is slightly higher than
for the full mass-limited sample while the scatter has 
decreased by only around 20-25 per cent.

While eliminating clusters with poor fits is an effective measure, 
it is not desirable for many applications where statistically-complete 
samples are required, most obviously when using cluster number counts 
to constrain cosmological parameters. Additionally, removing these 
objects could also introduce biases, e.g. in determining the 
distribution of cluster morphology or the range of thermodynamic 
profiles. Alternative, more optimal approaches may be possible such as 
measuring the profiles only along directions orthogonal to the merger 
axis, or using the azimuthal median profile rather than the mean. 
The latter, in particular, is a promising method to reduce the effect of 
shocks (or other discontinuities) that lead to local fluctuations and/or 
reduce gas clumping (e.g. \citealt{Zhuravleva2013,Eckert2015,Towler2023}).
We leave such a study to future work as it requires a detailed analysis 
of the SZ maps and the deprojection of these data to infer 
the underlying 3D profiles.

\subsection{Hydrostatic bias in different models}

\begin{figure}
\centering
\includegraphics[scale=0.5]{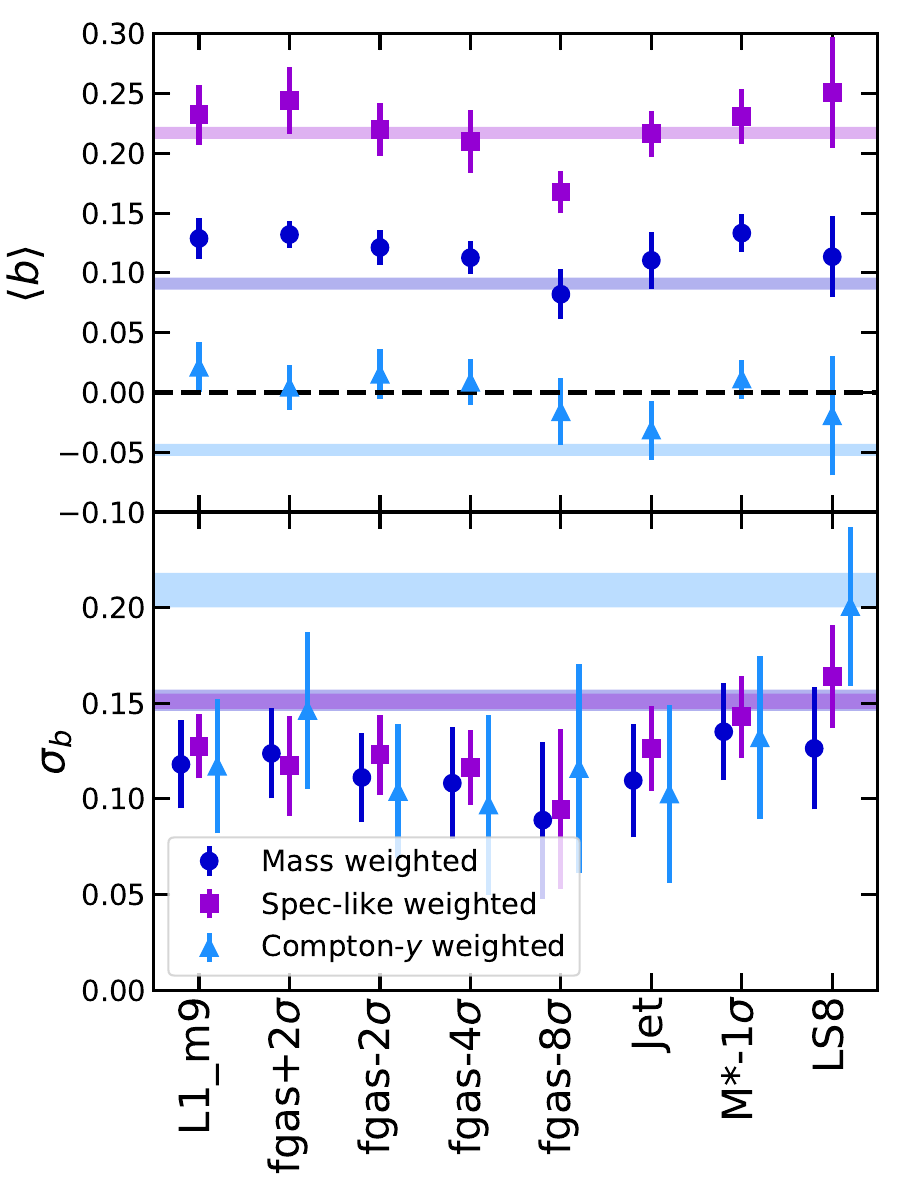}
\caption{
Median hydrostatic bias parameter, $\left< b \right>$ (top), and 
scatter, $\sigma_{b}$ (bottom), for massive 
($M_{\rm 500c}>7.5\times 10^{14}\,{\rm M}_{\odot}$) 
clusters in various L1 runs labelled along the x-axis. 
The symbols represent 
the different temperature weightings while error bars are the 1$\sigma$ 
uncertainties from bootstrap re-sampling 10,000 times. 
Horizontal bands represent results (and uncertainties) from the 
main L2p8\_m9 sample. The 
$\sigma_{b}$ bands for the mass-weighted and 
spectroscopic-like temperatures overlap.
}
\label{fig:hydrobias_comp}
\end{figure}

We compare the median bias, $\left< b \right>$, and scatter, $\sigma_{b}$, 
for various L1 runs in Fig.~\ref{fig:hydrobias_comp}. The former values 
are shown in the top panel along with the corresponding result from 
the L2p8\_m9 run as a horizontal band. 
For all models shown, we see the same separation in $\left< b \right>$ 
between the different temperature cases; using $T_{\rm sl}$ gives the 
largest bias and $T_{y}$ the smallest. For $T_{\rm sl}$, results for 
L1\_m9 are similar to L2p8\_m9 but the $T_{\rm m}$ and $T_{y}$ results 
increase with the latter now consistent with no bias. 
For a given temperature case, small differences are seen for the runs 
with varying gas fractions. 
As the gas fraction decreases (mainly due to stronger AGN feedback), the 
bias parameter goes down slightly. Runs with jet feedback, 
lower stellar masses or lower $S_{8}$ also produce similar bias 
parameters to the fiducial case.

The scatter in the bias parameter (bottom panel) is reasonably 
consistent between all runs and between different temperature 
weightings. However, the scatter for the $y$ weighted case is around 
a factor of two lower than for the larger L2p8\_m9 box. The $b$ 
distribution lacks the tail to negative $b$ values seen for L2p8\_m9 and 
also leads to a larger median value. The result cannot be explained by 
the larger box containing more massive objects as we saw earlier that the 
scatter did not vary with halo mass (Fig.~\ref{fig:bias_gof}). Instead, 
this suggests that the higher frequency of extreme clusters is a feature 
of the larger volume of the L2p8\_m9 run, better able to describe the 
non-Gaussian statistics of massive clusters. 

\subsection{Hydrostatic mass bias at higher redshift}

\begin{figure}
\centering
\includegraphics[scale=0.45]{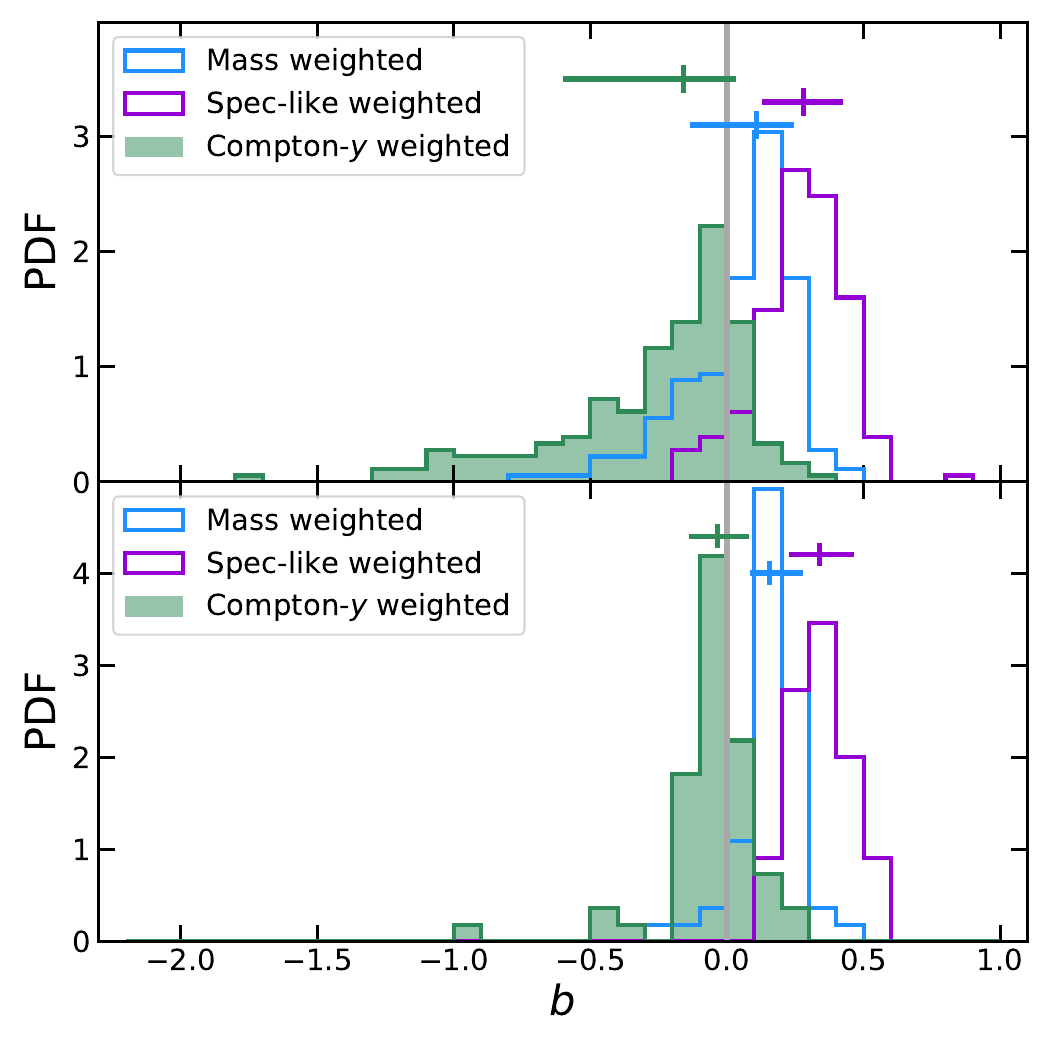}
\caption{
As in Fig.~\ref{fig:hse_bias} but showing the $b$ distributions 
for massive clusters at $z=0.5$.
}
\label{fig:hse_bias_z05}
\end{figure}

As stated above, our main results are presented at $z=0$ to 
maximise the sample size but many observed massive clusters 
will be located at intermediate redshifts (e.g. $z \sim 0.3-0.5$) as 
a result of the trade-off between the increase in volume and a lack of 
massive clusters existing at higher redshift (to illustrate this last point, 
we only find one cluster with $M_{\rm 500c}>10^{15}\,{\rm M}_{\odot}$ 
in L2p8\_m9 at $z=1$). We thus restrict our redshift study to $z=0.5$ 
where there are 181 objects above our mass limit.

Fig.~\ref{fig:hse_bias_z05} shows the equivalent set of results to 
Fig.~\ref{fig:hse_bias} but for $z=0.5$. We also provide median 
bias ($\left< b \right>$) and scatter ($\sigma_{b}$) values 
in Table~\ref{tab:bias_values}. For all 3 cases, 
the magnitude of the median bias increases at $z=0.5$ over $z=0$. 
However, the changes in the $y$-weighted bias are 
significantly larger than for the other two cases, 
with the median value around three times 
larger and the scatter increasing by around 50 per cent. 
These results are not unexpected since such massive 
clusters (selected above a fixed mass limit) are dynamically 
younger at higher redshift and are thus more likely to show 
signs of merger activity and shocks. We find similar trends 
between $b$ and $M_{\rm 500c}$, and $b$ and $\Delta(\mathcal{P})$, as 
at $z=0$. There is only a weak trend in $b$ with mass 
($r_{s}=(-0.06,0.05,-0.1)$ for the $T_{\rm m}, T_{\rm sl}, T_{y}$ cases, 
respectively) but a stronger (anti-)correlation between $b$ and 
$\Delta(\mathcal{P})$ ($r_{s}=(-0.4,-0.3,-0.5)$ for the same cases).
Again, restricting the sample to objects with $\Delta(\mathcal{P})<0.02$ 
(55 clusters), the bias for the $y$-weighted case reduces significantly 
with the scatter being consistent with the $z=0$ value at $1\sigma$.

\subsection{A closer look at the gas in cluster outskirts}

Our results have shown that, in line with previous studies, 
the SZ $y$-weighted temperatures are higher than the X-ray 
spectroscopic-like temperatures, 
particularly in the cluster outskirts. We have also shown that this 
leads to a smaller median hydrostatic mass bias around $R_{\rm 500c}$ 
than when the spectroscopic-like temperature is used, particularly when 
dynamically disturbed clusters with strong localised pressure 
fluctuations are excluded. It is tempting to think 
there is a {\it physical} reason for this reduction in bias as, 
away from shocks, gas with the highest thermal pressure ought to be 
the most {\it hydrostatic}. We investigate this here by 
comparing radial profiles for a few other gas properties 
using the same weightings as we did for the temperature. 

\begin{figure}
\centering
\includegraphics[scale=0.45]{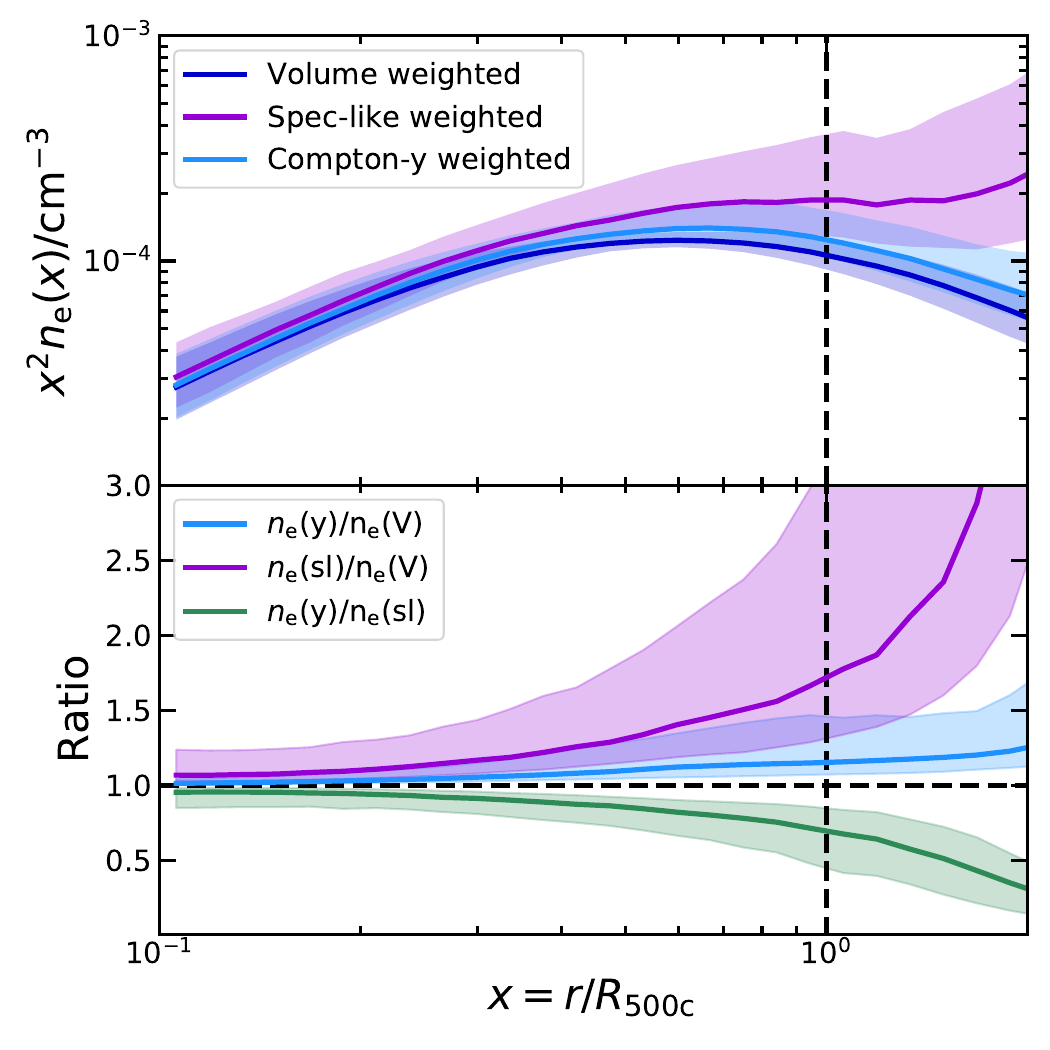}
\caption{Top: electron density profiles ($x^2 n_{\rm e}$) for clusters 
in L2p8\_m9 at $z=0$ for 3 different cases: volume weighted, 
spectroscopic-like weighted and $y$ weighted. 
Bottom: ratios between profiles with different weightings.}
\label{fig:prof3D_ne}
\end{figure}

\begin{figure*}
\centering
\includegraphics[scale=0.45]{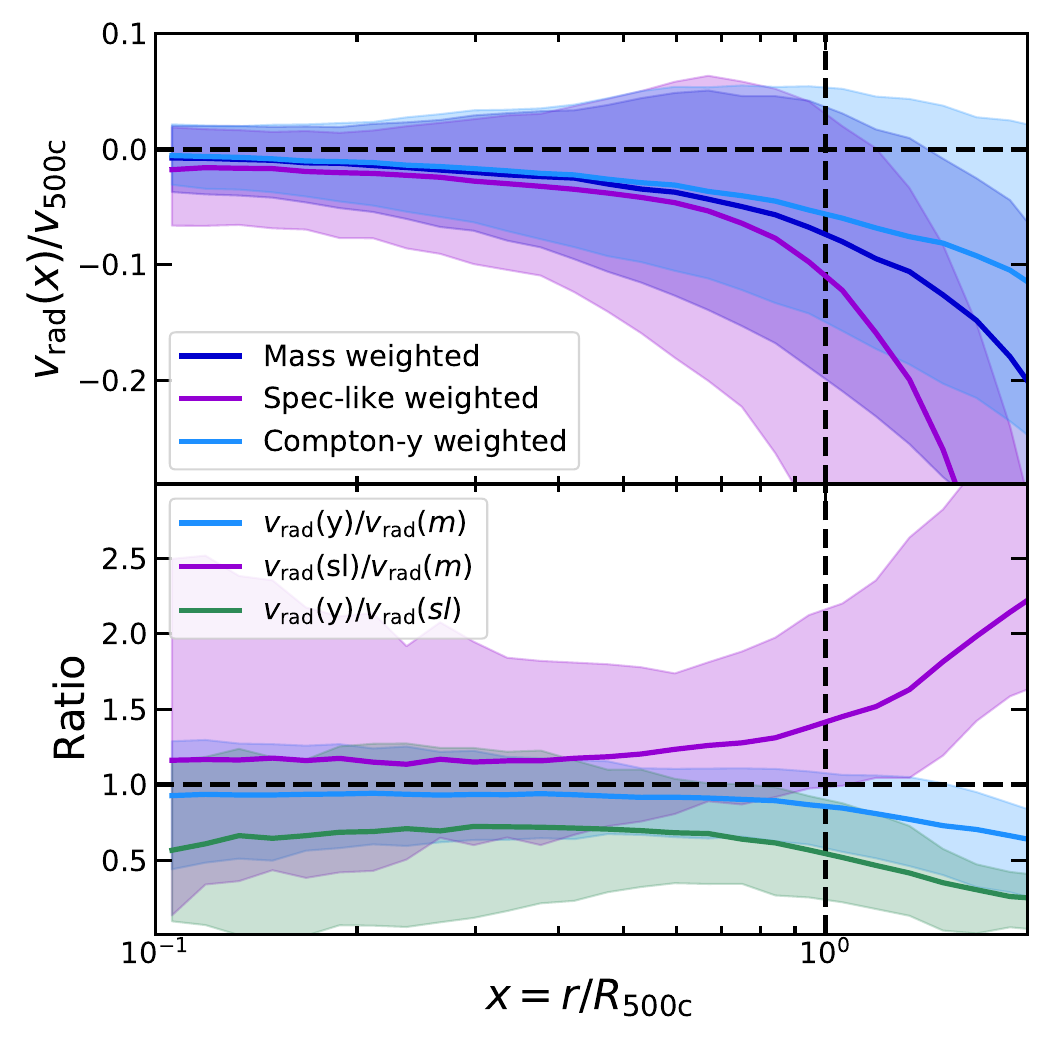}
\includegraphics[scale=0.45]{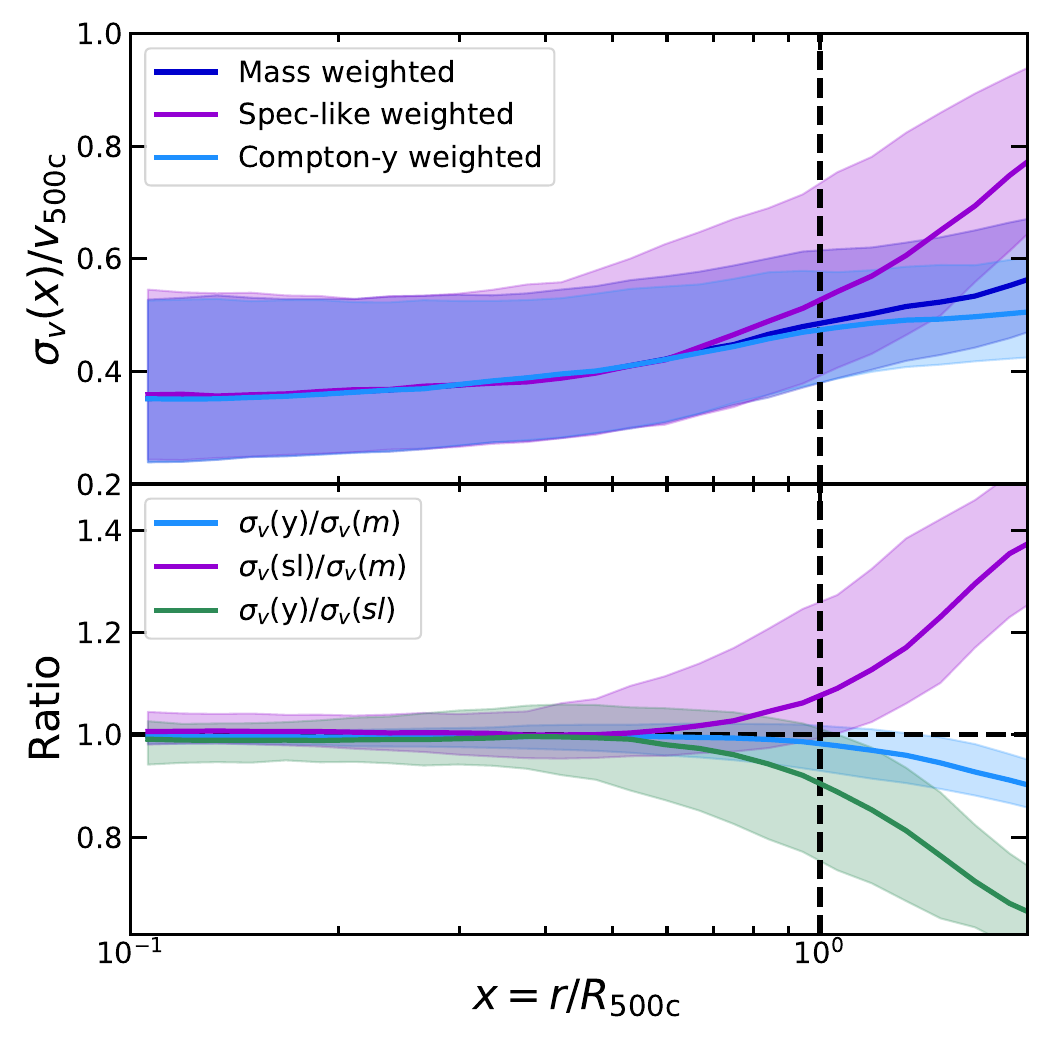}
\caption{Left: radial velocity profiles for the massive clusters 
in L2p8\_m9 at $z=0$ for 3 different cases: mass weighted, 
spectroscopic-like weighted and $y$ weighted. 
Right: 3D velocity dispersion profiles for the same cases.
Ratios of the profiles are shown in the lower panels (the absolute 
value is taken for the radial velocity case).}
\label{fig:prof3D_vradsigmav}
\end{figure*}

Fig.~\ref{fig:prof3D_ne} shows the electron density 
($x^2n_{\rm e}$) profiles for 
three different weightings. The first case is the volume-weighted 
electron density which has the smallest cluster-to-cluster scatter. 
We also show the case that uses spectroscopic-like weighting; while 
not an observable, it informs us of the typical electron density 
of gas that contributes most to $T_{\rm sl}$. At $x>0.6$, we see 
that $x^{2}n_{\rm e}$ is approximately constant and the ratio between 
this density with the volume-weighted case increases rapidly with 
radius. This is a result of the ICM becoming increasingly clumpy 
at larger radii, with the denser gas also tending to be cooler 
(both increasing the spectroscopic-like weight).  
On the other hand, the $y$-weighted profile traces the volume-weighted 
profile more closely, to within 25 per cent or so, out to $x=2$.
Thus, the gas in the outskirts with the highest pressure (and thus the 
largest $y$ weighting) has lower density and higher temperature 
than the gas with the highest spectroscopic-like weighting. On 
larger scales (not shown), we find that even the $y$-weighted 
density becomes significantly higher than the volume-weighted case, 
particularly around $5R_{\rm 500c}$, where we expect the accretion 
shock to be located; see e.g. \cite{Aung2021}.

This result suggests that the high-pressure gas, traced by the SZ 
effect, may be more {\it hydrostatic} in the outskirts, on average. 
To test this, 
we also consider the weighted hot gas radial velocity and 3D velocity 
dispersion profiles as both should be reduced for the $y$-weighted gas, 
relative to the other weightings.
Fig.~\ref{fig:prof3D_vradsigmav} confirms this, where 
we have scaled the velocities to the circular velocity, 
$v_{\rm 500c}$, of each halo 
at $R_{\rm 500c}$. In the left panel, the median radial velocities 
are always negative, as expected for infalling gas. In all cases, these 
values are small ($<$10 per cent of $v_{\rm 500c}$) within 
$R_{\rm 500c}$. At larger radii, the radial velocity magnitudes are 
larger but the $y$-weighted median remains within 10 per cent of 
$v_{\rm 500c}$ or so, at $x=2$. On the other hand, the 
spectroscopic-like-weighted gas has a significantly larger 
radial velocity magnitude, over twice that of the mass-weighted 
case (and around 4 times that of the $y$-weighted case) at $x=2$.
Similarly, in the right panels, the 3D velocity dispersions are 
similar to each other within $R_{\rm 500c}$, with $\sigma_{v}$ 
increasing with radius, signifying an increase in non-thermal pressure. 
In the outskirts, the median values for the three weightings diverge, with 
the $y$-weighted value being the smallest and the spectroscopic-like-weighted 
value the largest (the former is around two thirds of the latter at $x=2$).

\section{Summary and Conclusions}
\label{sec:conclusions}

In this paper, we have used the large-volume FLAMINGO cosmological 
simulations \citep{Schaye2023,Kugel2023}
to study how using the Compton $y$-weighted temperature, 
an observable that can be extracted from the relativistic 
SZ spectral distortion in massive clusters, impacts upon 
measurements of the hydrostatic mass bias. We also study how 
this temperature compares with the mass-weighted and 
spectroscopic-like temperatures, the latter being a proxy 
for the X-ray spectroscopic temperature, both within clusters 
and in their outskirts. The FLAMINGO simulations have been 
particularly beneficial as they have allowed us to 
(a) select, for the first time, a large ($\sim 10^{3}$ objects) 
mass-complete sample of 
massive ($M_{\rm 500c}>7.5\times 10^{14}\,{\rm M}_{\odot}$) clusters 
from a single hydrodynamical simulation; 
and (b) study the impact of varying the cosmological and subgrid 
(feedback) models relative to the fiducial case that is calibrated 
to observations. The fiducial model has also previously been shown to 
produce results in good agreement with observed 
X-ray cluster scaling relations and radial profiles 
\citep{Braspenning2023}. Here, we summarise our main results:
\begin{itemize}

\item Compton-$y$-weighted temperatures are higher than mass-weighted and 
spectroscopic-like temperatures at fixed mass, in line with 
previous work, and evolve self-similarly with redshift out to at least $z=1$ 
(Fig.~\ref{fig:tmrel_z_llr}). These temperatures are higher when 
clusters contain lower gas fractions (mainly as a result of stronger 
AGN feedback). The temperature scatter at fixed mass is insensitive to 
these changes in high mass clusters but increases with feedback 
strength in groups (Fig.~\ref{fig:tmrel_agn_llr}).

\item Pressure profiles are generally well characterized by the 
GNFW model for massive clusters (Fig.~\ref{fig:prof3D_Pe}). 
Cluster-to-cluster scatter is minimal at $r \approx 0.6 R_{\rm 500c}$ 
and maximal at $r \approx 3 R_{\rm 500c}$. The profiles are self-similar 
at these scales to good approximation (Fig.~\ref{fig:prof3D_Pe_M500c}). 
Models with stronger AGN feedback have 
slightly lower pressure in the core and higher pressure in the outskirts, 
as a result of more gas being ejected (Fig.~\ref{fig:pprof3d_comp}).

\item $y$-weighted temperatures are similar to spectroscopic-like 
temperatures around the edge of the core ($r \approx 0.1 R_{\rm 500c}$) 
but diverge at larger radii and are around twice as high in the 
outskirts ($r \approx 2R_{\rm 500c}$; Fig.~\ref{fig:prof3D_T}).
At $R_{\rm 500c}$, appropriate for mass estimates, $T_{y}$ is 
around 25 per cent higher than $T_{\rm sl}$ and increases gradually 
with mass (Fig.~\ref{fig:tprof3d_ratio}). We find an increase in 
$T_{y}/T_{\rm sl}$ with radius that is insensitive to variations 
in baryonic physics and cosmology (Fig.~\ref{fig:tprof3d_comp}), 
making it a robust prediction that can be tested with X-ray and 
SZ observational data e.g. by stacking clusters to measure the signal 
in the outskirts.

\item Hydrostatic masses with $y$-weighted temperatures are less 
biased, on average, than when using spectroscopic-like temperatures 
but have larger scatter (Fig.~\ref{fig:hse_bias}). The latter 
is due to clusters with large pressure and $T_{y}$ fluctuations 
close to $R_{\rm 500c}$ (Fig.~\ref{fig:profiles_goodbad}) that, 
on inspection, are associated with merger activity 
(Fig.~\ref{fig:temp_maps}). Such clusters tend to have poorly fitting 
GNFW models (Fig.~\ref{fig:bias_gof}) and when only clusters with 
good fits are considered, the scatter reduces considerably. At 
intermediate redshift ($z=0.5$) the median bias is around three times 
larger as a result of massive clusters being dynamically younger 
(Fig.~\ref{fig:hse_bias_z05}) but is consistent with the $z=0$ result 
when poorly fitted clusters are removed. 

\item The median bias and scatter are similar for all 
models run with a 1 Gpc box and varying baryonic physics and 
cosmology (Fig.~\ref{fig:hydrobias_comp}) but show significant 
deviations from the main results that used a larger (2.8 Gpc) box. 
The former do not contain as many extreme, merging objects so is likely 
a statistical effect caused by the non-Gaussian nature of the density 
field in relation to rare, massive objects. It is therefore evident 
from this example that caution should be applied when using cosmological 
simulations with modest box sizes to calculate probabilities of rare 
events like merging massive clusters (extreme value statistics). 

\item Focusing on the gas in cluster outskirts ($1<r/R_{\rm 500c}<2$), 
Compton-$y$ (pressure) weighting yields lower electron density 
(Fig.~\ref{fig:prof3D_ne}), radial velocity and velocity dispersion 
(Fig.~\ref{fig:prof3D_vradsigmav}) than X-ray (spectroscopic-like) 
weighting, as well as higher temperatures. These results suggest the 
SZ temperature is a more sensitive tracer of gas that is 
smoother and more {\it hydrostatic} than the X-ray temperature, 
which is affected by cooler, clumpier gas. 
\end{itemize}

In conclusion, our study shows that the relativistic SZ effect in 
clusters is an important method for independently measuring the ICM 
temperature. Firstly, there is the prospect of measuring cluster 
masses with SZ data only; our results show these masses to be 
unbiased if pressure fluctuations associated with mergers/shocks can 
be accounted for. Secondly, our work shows that comparing SZ and 
X-ray temperatures in cluster outskirts ought to be informative 
for probing the gas structure. The ability to measure the temperature 
of the ICM with SZ observations may also be useful as an independent 
calibration measure, as temperature measurements from different 
X-ray telescopes differ at the level of 10-20 per cent 
\citep[e.g.][]{Schellenberger2015,Wallbank2022}. 

Existing attempts of measuring SZ temperatures are still quite 
rare but future telescopes with millimetre/sub-millimetre capability 
are including the relativistic SZ effect in their science case, e.g. 
FYST/CCAT-prime \citep{Stacey2018}; 
AtLAST \citep{Ramasawmy2022,DiMascolo2024}. With such instruments, it should 
be possible to measure relativistic temperatures at different 
scales in a reasonable amount of observing time 
\citep{Perrott2023} and start to test the above predictions.
On the theoretical side, we will also require more realistic, mock 
X-ray and SZ data from simulations to test whether our predictions 
hold under more realistic observational conditions.

\section*{Acknowledgements}
This work used the DiRAC@Durham facility managed by the Institute for 
Computational Cosmology on behalf of the STFC DiRAC HPC Facility 
(www.dirac.ac.uk). The equipment was funded by BEIS capital funding via STFC 
capital grants ST/K00042X/1, ST/P002293/1, ST/R002371/1 and ST/S002502/1, 
Durham University and STFC operations grant ST/R000832/1. DiRAC is part of 
the National e-Infrastructure. This work is partly funded by research 
programme Athena 183.034.002 from the Dutch Research Council (NWO).
JC was supported by the ERC Consolidator Grant {\it CMBSPEC} (No.~725456) and the Royal Society as a Royal Society University Research Fellow at the University of Manchester, UK (No.~URF/R/191023).

\section*{Data Availability}

The data used in the plots can be made available on reasonable 
request to the corresponding author. The FLAMINGO data will 
eventually be made publicly available when it is practically 
feasible to do so. In the meantime, anyone interested in using 
the data can contact the corresponding author.



\bibliographystyle{mnras}
\bibliography{SZmass} 



\appendix

\section{Higher order terms}
\label{sec:res_higher}

Here, we briefly look at the radial profiles of the functions used 
for the higher (2nd, 3rd and 4th) order terms in 
the Taylor expansion of the relativistic spectral distortion 
$f(\nu,T_{\rm e})$ about $T_{y}$ \citep{Chluba2013}. 
As discussed in Section~\ref{sec:relSZ}, this choice of pivot 
temperature means that the linear term vanishes and 
the $k$th order terms are proportional to the following temperature 
moments
\begin{equation}
    \mathcal{F}_{k} = \left< (T_{\rm e}-T_{y})^{k} \right>, 
\label{eqn:higher}
\end{equation}
where $\left< \right>$ corresponds to the y-weighted average 
used in equation~\ref{eqn:Ty}. For $k=2,3\, \& \, 4$, we can write 
the respective functions as 
\begin{eqnarray}
\mathcal{F}_{2} \equiv \sigma_{y}^{2} &=& 
\left< T_{\rm e} \right>^{2}-T_{y}^{2} 
\nonumber \\
\mathcal{F}_{3} \equiv \rho_{y}^{3} &=& 
\left< T_{\rm e}^{3} \right> + 2T_{y}^{3} - 3T_{y}\left< T_{\rm e}\right>^{2} 
\nonumber \\
\mathcal{F}_{4} \equiv \kappa_{y}^{4} &=& 
\left< T_{\rm e}^{4} \right> - 3T_{y}^{4} + 6T_{y}^{2}\left< T_{\rm e}\right>^{2} - 4T_{y}\left< T_{\rm e} \right>^{3},
\end{eqnarray}
where $\sigma_{y},\rho_{y},\kappa_{y}$ all have dimensions of temperature.
(Note that these functions will be related to the variance, 
skewness and kurtosis of the temperature distribution at a given 
radius, respectively.) Thus, to fourth order, we have
\begin{equation}
    \int \! \tilde{y} f(\nu,T_{\rm e}) {\rm d}V \approx 
    Y \left[ f(\nu,T_{y}) + 
    \tfrac{1}{2}\partial^{2}\! f \sigma_{y}^{2} + 
    \tfrac{1}{6}\partial^{3}\! f \rho_{y}^{3} + 
    \tfrac{1}{24}\partial^{4}\! f \kappa_{y}^{4}
    \right].
\end{equation}

\begin{figure}
\centering
\includegraphics[scale=0.45]{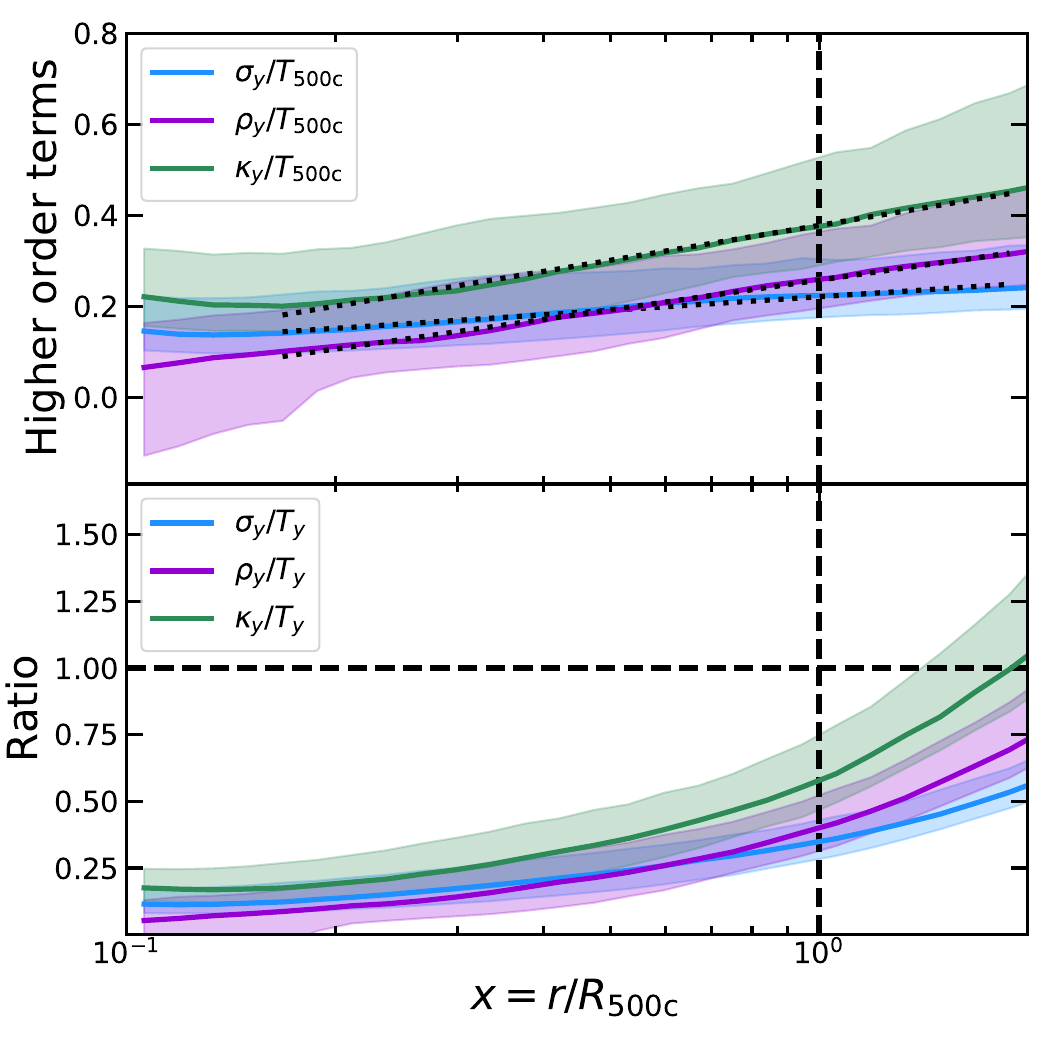}
\caption{Radial dependence of the higher order terms in the 
relativistic SZ expansion (see text for details). Top: results are scaled 
to $T_{\rm 500c}$ for each cluster. The dotted lines are least-squares 
fits to the median profiles over the radial range $0.15<r/R_{\rm 500c}<2$.
Bottom: ratio of each higher order term to $T_{y}$.}
\label{fig:prof3D_higher}
\end{figure}

\begin{table}
	\centering
	\caption{Values for the intercept, $A$, and slope, $B$, for 
    linear least-squares fits to the variation in 
    higher-order terms, relative to $T_{\rm 500c}$, with radius.}
	\label{tab:higher_order_fits}
	\begin{tabular}{lll} 
		\hline
        Term & $A$ & $B$\\
		\hline
        $\sigma_{y}/T_{\rm 500c}$ & $0.22$ & $0.10$\\
        $\rho_{y}/T_{\rm 500c}$ & $0.26$ & $0.22$\\
        $\kappa_{y}/T_{\rm 500c}$ & $0.38$ & $0.25$\\
		\hline
	\end{tabular}
\end{table}

We evaluate the temperature moments, 
$\sigma_{y},\rho_{y},\kappa_{y}$, as a function of radius and 
show the results in Fig.~\ref{fig:prof3D_higher}. 
The top panel shows that, unlike $T_{y}$, these higher order 
functions increase with radius, i.e. the ($y$-weighted) 
temperature distribution is becoming increasingly broad and 
asymmetric on larger scales. 
Relative to $T_{y}$, these terms are small (around 
10 per cent in the core, $x=0.1$) but become significant 
in the outskirts ($x=1-2$), as shown in the bottom panel.
A similar result was found by \cite{Lee2020} for the 
temperature scatter, $\sigma_{y}$, using the BAHAMAS+MACSIS 
simulations \citep{McCarthy2017,Barnes2017}. 
We provide a simple linear least-squares fit to the radial profiles in the 
form $Y = A + B\log_{10}(r/R_{\rm 500c})$ where 
$Y=\{ \sigma_{y},\rho_{y},\kappa_{y} \}/T_{\rm 500c}$ and the fit 
is performed over the radial range $0.15<r/R_{\rm 500c}<2$. 
Values for $A$ and $B$ are given in Table~\ref{tab:higher_order_fits} with 
the fits shown as dotted lines in the top panel of 
Fig.~\ref{fig:prof3D_higher}.

\begin{figure}
\centering
\includegraphics[scale=0.5]{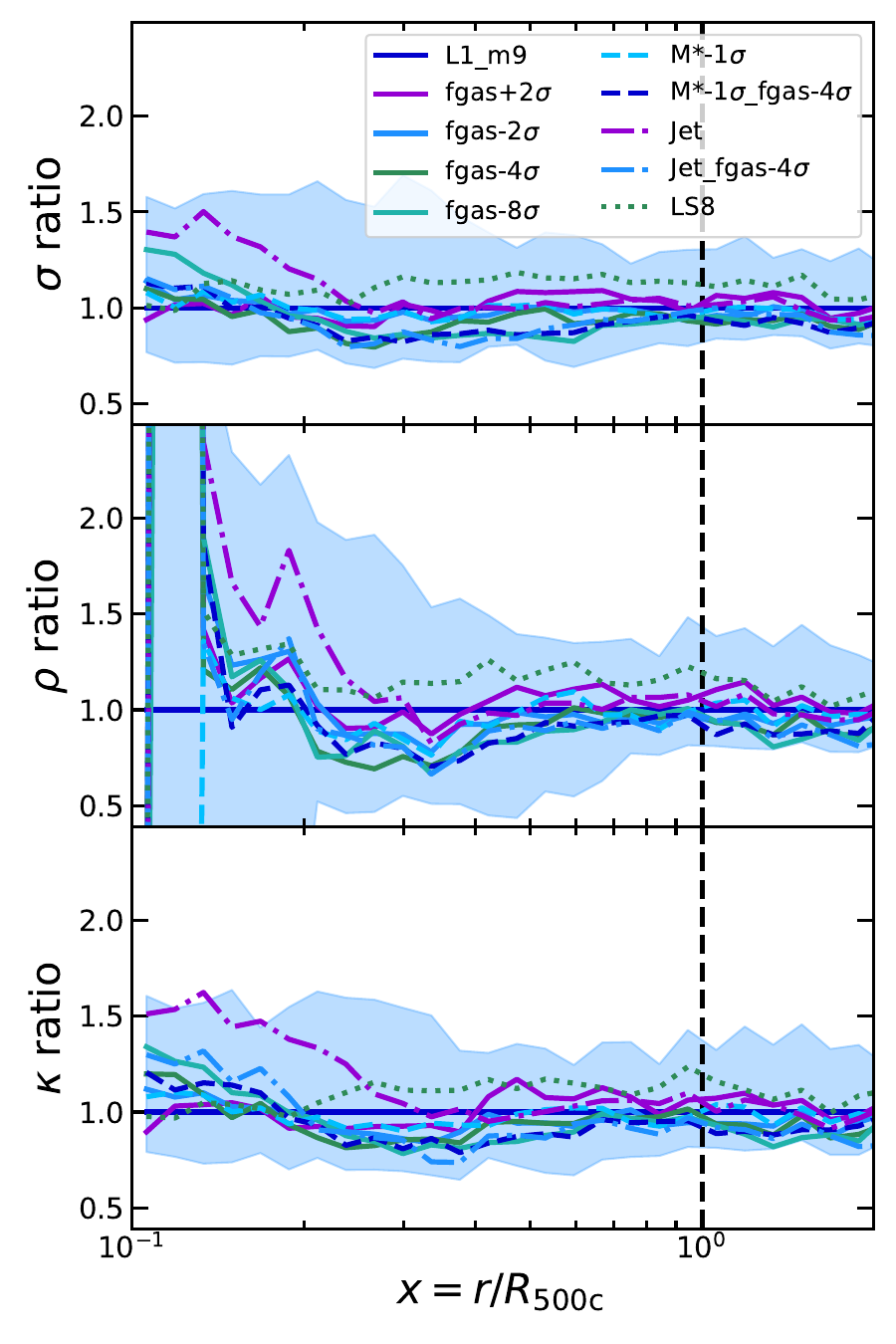}
\caption{Ratio of median higher order profiles in the varying L1 models 
to the fiducial L1\_m9 case}
\label{fig:prof3D_higher_comp}
\end{figure}

We also compare the radial profiles of the higher order terms 
between models in Fig.~\ref{fig:prof3D_higher_comp}. As with 
the pressure and temperature profiles, the differences  
between models are typically small, although the kinetic 
AGN feedback Jet model produces larger temperature scatter 
(and higher-order effects) in the central region ($x<0.2$ or so).

Note that each higher order ($k$) term  
would also need to be weighted by $\partial^{k}f/k!$ 
when calculating the full relativistic signal at a given frequency. 
This can be achieved using numerical codes  
like {\sc szpack} \cite{Chluba2012}, something we will address 
in future work.

\section{Results at varying resolution}
\label{sec:res_study}

\begin{table}
	\centering
	\caption{Fiducial runs with varying box-size and resolution. 
 Column~1 gives the run label; 2 and 3 the baryon and particle 
 numbers respectively; 4 the comoving box-size and 5 the gas particle mass.
 Note that runs at varying resolution are calibrated separately to the 
 observational data.}
	\label{tab:flamingo_parameters}
	\begin{tabular}{lrrll} 
		\hline
        Label & $N_{\rm b}$ & $N_{\nu}$ & $L$ & $m_{\rm gas}$\\
        & & & (cGpc) & ${\rm M}_{\odot}$\\
		\hline
        L2p8\_m9 & 5040$^{3}$ & 2800$^3$ & 2.8 & $1.07\times 10^{9}$\\
        L1\_m8 & 3600$^{3}$ & 2000$^3$ & 1   & $1.34\times 10^{8}$\\
        L1\_m9 & 1800$^{3}$ & 1000$^3$ & 1   & $1.07\times 10^{9}$\\
        L1\_m10& 900$^{3}$  & 500$^3$  & 1   & $8.56\times 10^{9}$\\
		\hline
	\end{tabular}
\end{table}

The FLAMINGO simulations include L1 runs at varying 
resolution; here we compare some of our key results 
for the runs with gas particle masses 8 times higher (L1\_m10) and 
lower (L1\_m8) than the fiducial case (L1\_m9); see 
Table~\ref{tab:flamingo_parameters}. 
Note these runs were calibrated separately (to the same 
observables) so their 
sub-grid model parameters vary (see \citealt{Kugel2023} and 
\citealt{Schaye2023} for details). As discussed in 
\cite{Schaye2015}, this can be classed as a {\it weak} 
convergence study. 

\begin{figure*}
\centering
\includegraphics[scale=0.6]{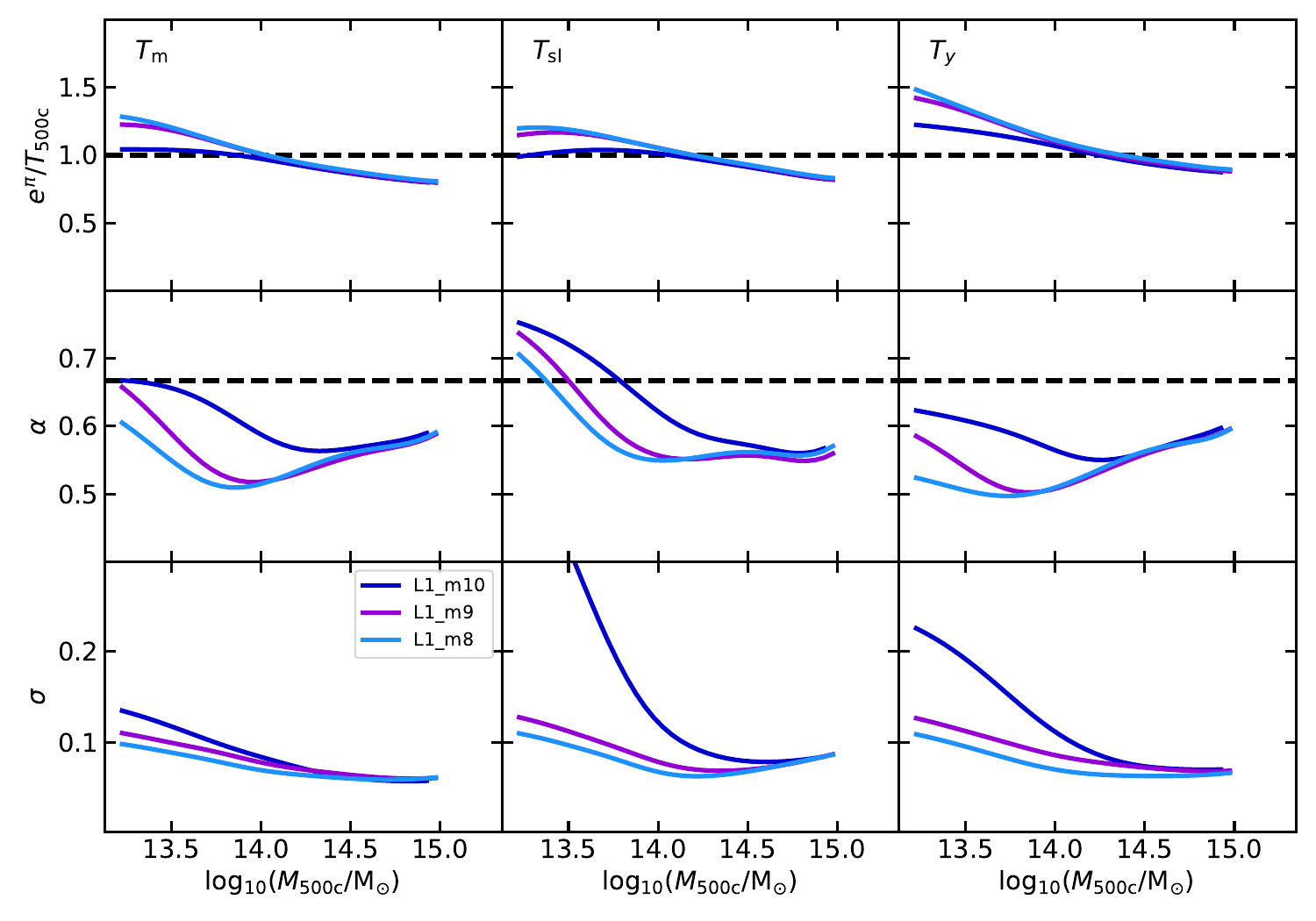}
\caption{Temperature-mass relations for the L1 box with 
varying mass resolution 
(L1\_m10, L1\_m9, L1\_m8 are low, standard, and high resolution 
runs, respectively).  
Details are as in Fig.~\ref{fig:tmrel_z_llr}.}
\label{fig:tmrel_res_llr}
\end{figure*}

Fig.~\ref{fig:tmrel_res_llr} shows the LLR parameters 
for the temperature-mass relations at $z=0$. The normalisation 
parameter (top panel) is converged for all 3 runs on cluster 
scales ($M_{\rm 500c}>10^{14}\,{\rm M}_{\odot}$), whereas on group 
scales, the low-resolution L1\_m10 run under-predicts the temperature 
by up to 30 per cent or so (this is likely due to stellar feedback 
being switched off in this run which also predicts larger gas fractions 
in lower-mass groups which are below the calibration scale; 
see Fig.~10 in \citealt{Schaye2023}). 
The slope and scatter (middle and bottom panels) show larger differences 
at low mass, particularly between the the L1\_m10 and L1\_m9 models.
Both the slope and scatter are smaller 
at higher resolution. The L1\_m9 and L1\_m8 runs show better agreement, 
with the minimum slope occuring around the same halo mass.
All 3 runs come into good agreement at 
$M_{\rm 500c}>10^{14.3}\,{\rm M}_{\odot}$.

\begin{figure}
\centering
\includegraphics[scale=0.45]{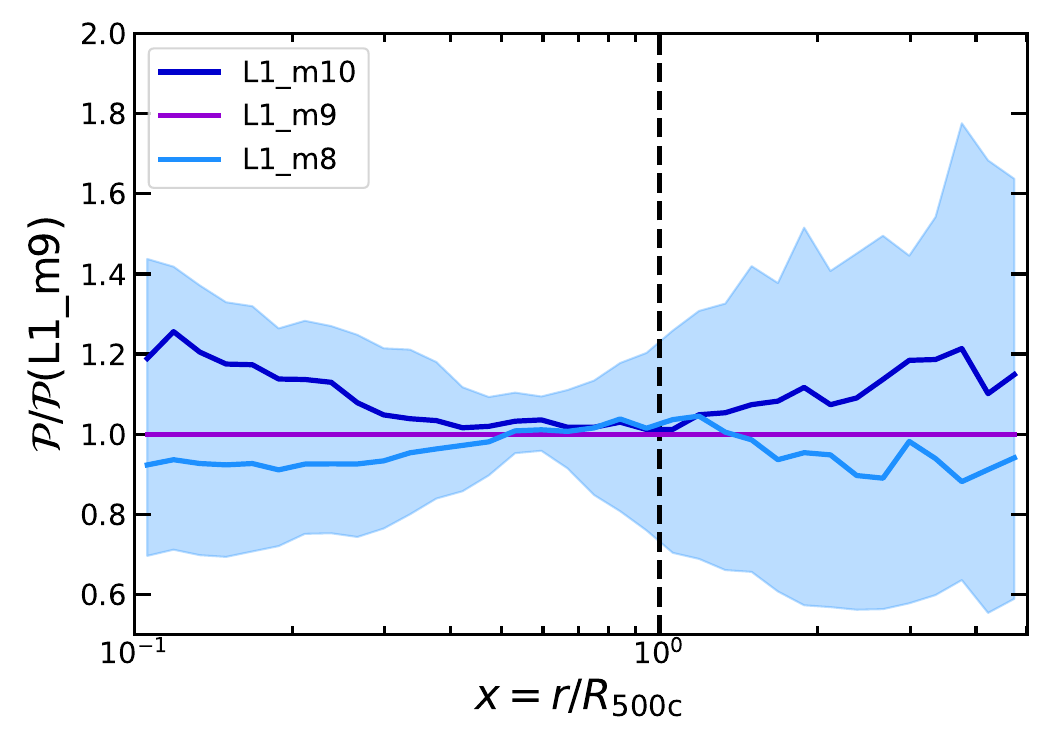}
\caption{As in Fig.~\ref{fig:pprof3d_comp} 
but comparing the median pressure profile for 
each resolution level to the fiducial case.}
\label{fig:pprof3d_comp_res}
\end{figure}

\begin{figure}
\centering
\includegraphics[scale=0.45]{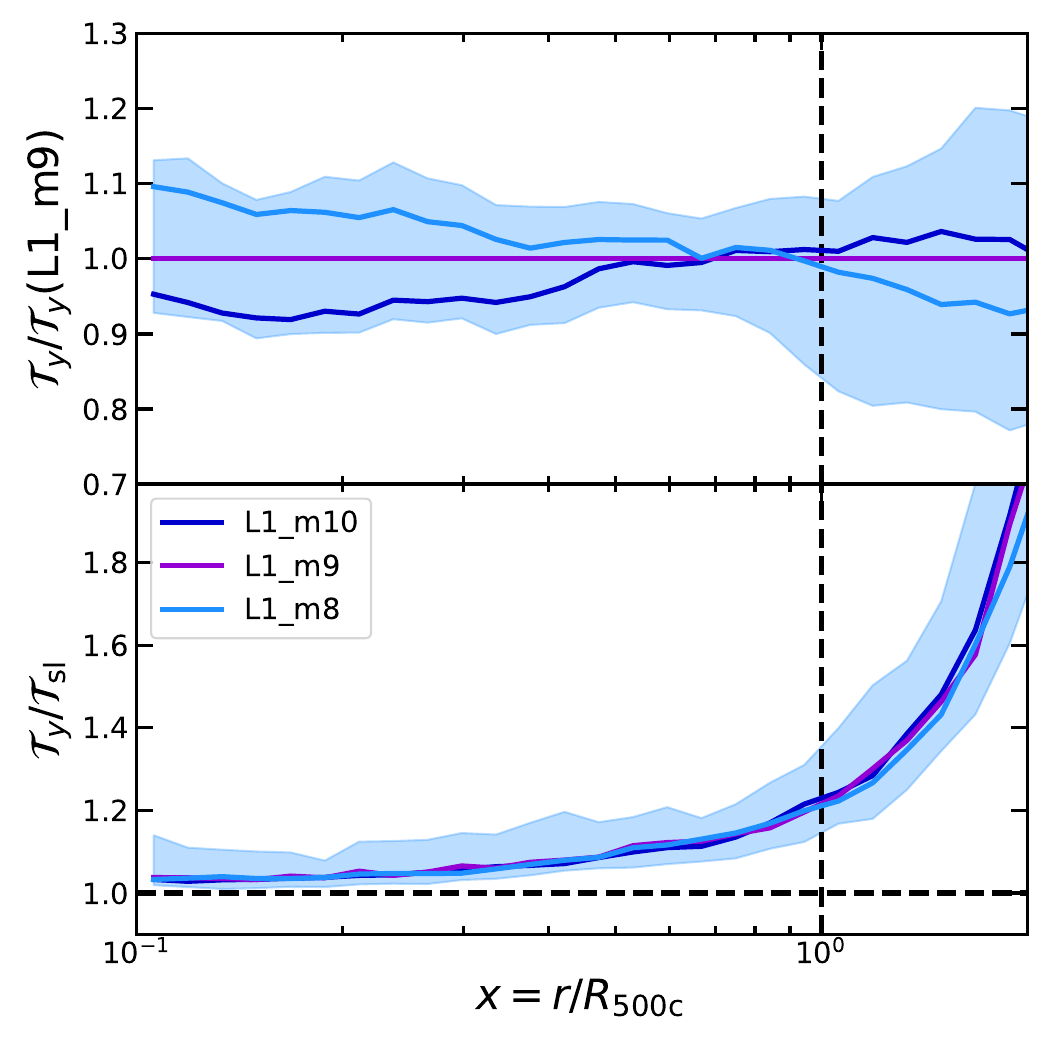}
\caption{As in Fig.~\ref{fig:tprof3d_comp}
but comparing the median temperature profile for 
each resolution level to the fiducial case.}
\label{fig:tprof3d_comp_res}
\end{figure}

Fig.~\ref{fig:pprof3d_comp_res} compares the median 
pressure profiles for the massive cluster sample 
($M_{\rm 500c}>7.5\times 10^{14}\,{\rm M}_{\odot}$) 
at the three resolution levels. As in 
Fig.~\ref{fig:pprof3d_comp}, we plot the profile relative to 
the fiducial L1\_m9 case. In general, the median profiles are very similar, 
with deviations only occurring in the core and far outskirts.
The lower resolution L1\_m10 run produces clusters 
with slightly higher (20 per cent) pressures in the 
core and far outskirts, while the higher resolution  
L1\_m8 run only differs by up to 5 per cent or so. 
Importantly, for this study, the pressure profiles are 
well converged around $R_{\rm 500c}$, where we estimate 
cluster masses.

\begin{figure}
\centering
\includegraphics[scale=0.5]{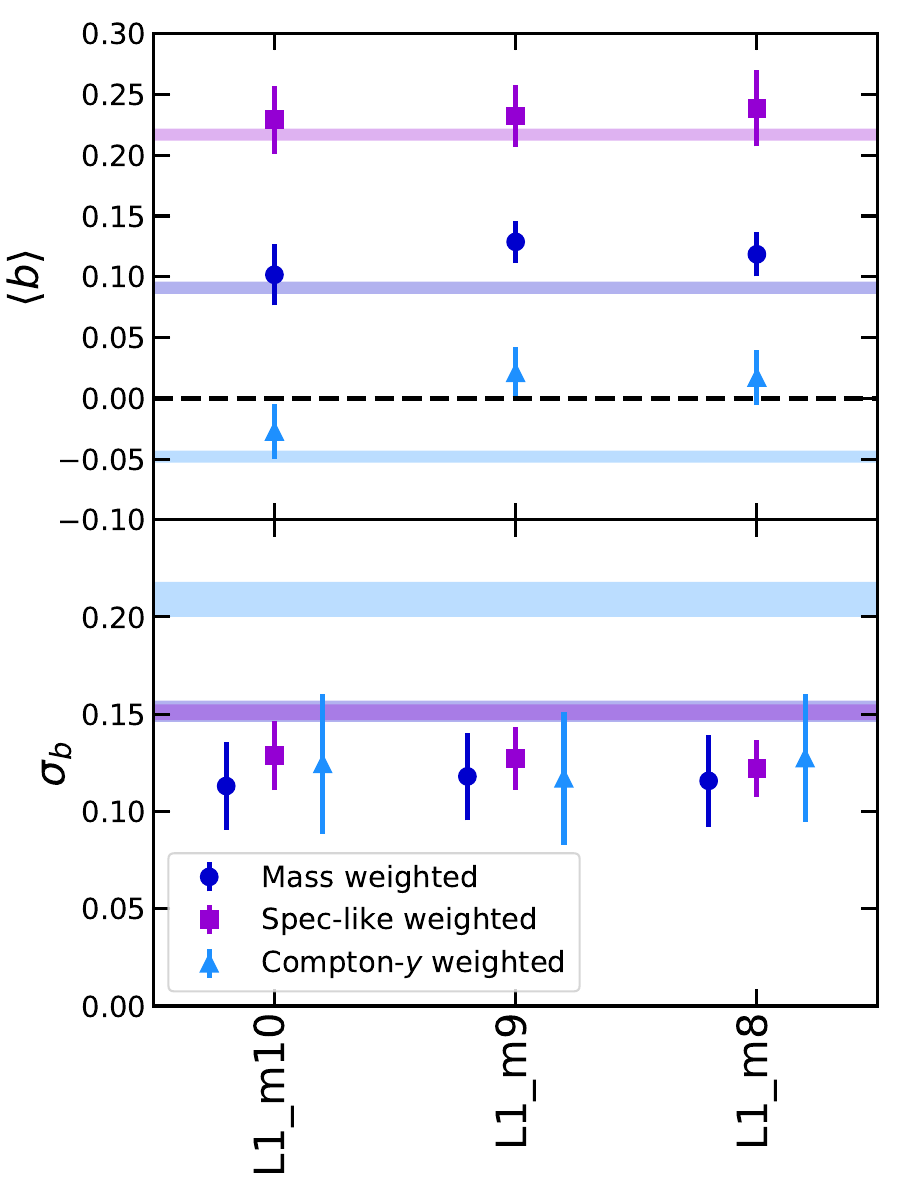}
\caption{
As in Fig.~\ref{fig:hydrobias_comp} but comparing the 
bias parameter for the runs with different resolution.
}
\label{fig:hydrobias_comp_res}
\end{figure}

We also compare $y$-weighted temperature profiles in Fig.~\ref{fig:tprof3d_comp_res}. Here, the 
deviations are smaller (within 10 per cent) with a higher core temperature at higher resolution. 
The increasing $T_{y}/T_{\rm sl}$ ratio with radius is identical in all 3 resolution cases, 
providing further support to the robustness of this result. 

Finally, we compare median hydrostatic bias parameters, $\left< b \right>$, and 
scatter, $\sigma_{b}$, for the runs with varying resolution in Fig.~\ref{fig:hydrobias_comp_res}.
Results for all three runs are consistent with each other, with the $y$-weighted masses being 
unbiased, on average.


\bsp	
\label{lastpage}
\end{document}